\documentclass[12pt,a4paper, fleqn]{article}

\input{preamble}

\author{\textsc{Egshiglen Batbayar}\thanks{Department of Economics, University of Bonn, Adenauerallee 24-26, 53113 Bonn, Germany, e-mail:
\url{ebatbayar@uni-bonn.de}}
\\{\small \textit{University of Bonn}}
\and\textsc{Christoph Breunig}\thanks{Department of Economics, University of Bonn, Adenauerallee 24-26, 53113 Bonn, Germany, e-mail: \url{cbreunig@uni-bonn.de}}
\\{\small \textit{University of Bonn}}
  \and\textsc{Peter Haan}\thanks{DIW Berlin  and Freie Universit\"at Berlin, Mohrenstr. 58, 10117 Berlin, Germany, e-mail: \url{ phaan@diw.de}}\\
  {\small \textit{DIW Berlin, Freie Universit\"at Berlin}}
   \and{\textsc{Boryana Ilieva}\thanks{DIW Berlin and European Central Bank, Sonnemannstrasse 20, 60314 Frankfurt am Main, Germany, e-mail: \url{boryana.ilieva@ecb.europa.eu}}}
  \\
 {\small \textit{DIW Berlin, ECB} }
}

\title{Quantile Selection in the Gender Pay Gap\thanks{We thank Martin Biewen, Richard Blundell, Joachim Freyberger, Björn Höppner, Maximilian Schaller, Jan Scherer, Arne Uhlendorff and numerous seminar and conference participants for helpful comments and illuminating discussions. Christoph Breunig gratefully acknowledges the support of the Deutsche Forschungsgemeinschaft (DFG, German Research Foundation) under Germany’s Excellence Strategy – EXC-2047/1 – 390685813. Peter Haan acknowledges the support by the DFG through CRC TRR 190 (project number 280092119), FOR 5675 (project number 518302089) and HA 5526/7-1.}
}

\begin{document}
   \maketitle
\begin{abstract}
{\small \noindent We propose a new approach to estimate selection-corrected quantiles of the gender wage gap. Our method employs instrumental variables that explain variation in the latent variable but, conditional on the latent process, do not directly affect selection. We provide semiparametric identification of the quantile parameters without imposing parametric restrictions on the selection probability, derive the asymptotic distribution of the proposed estimator based on constrained selection probability weighting, and demonstrate how the approach applies to the Roy model of labor supply. Using German administrative data, we analyze the distribution of the gender gap in full-time earnings. We find pronounced positive selection among women at the lower end, especially those with less education, which widens the gender gap in this segment, and strong positive selection among highly educated men at the top, which narrows the gender wage gap at upper quantiles.
}
\end{abstract}
{\small \begin{tabbing}
 \> parametric inference, gender wage gap, wage inequality\\[.2ex]
\end{tabbing}}

\clearpage

\section{Introduction}

In all countries, men continue to earn higher wages than women. Crucially, the gender wage gap spans the entire wage distribution and affects all educational groups (e.g., \cite{Goldin_14}, \cite{blau2017gender}, \cite{OlPaPa_2024}). 
To meaningfully quantify and analyze this gap, it is essential to account not only for gender differences in observable characteristics such as education and labor market experience, but also differences in selection into employment. The impact of selection on the gender wage gap is ambiguous and depends on the sign and magnitude of gender-specific selection patterns (see, e.g., \cite{mulligan2008selection}, \cite{arellano2017}, or \cite{Blau_etal_2021}). Understanding why men consistently out-earn women and designing policies to address this disparity requires tackling the challenging task of identifying and quantifying the role of employment selection in shaping the wage distribution, and of deriving a selection-corrected measure of the gender wage gap distribution.

In this paper, we propose a new strategy to derive the selection-corrected wage distribution and to quantify the full distribution of the gender wage gap.
Identification relies on exogenous variation that affects latent wages but, conditional on the latent wages and other observed variables, provides no additional information on the selection mechanism.
In the context of a Roy model \citep{Roy_51}, as applied to labor supply decisions by \cite{gronau1974} and \cite{heckman1974}, where individuals choose to work if their potential wage exceeds their reservation wage, we show that our exogeneity requirements have the interpretation of a \emph{rank invariance} condition: Conditional on potential wages, the instrument does not provide additional information about an individual’s position in the reservation wage distribution.

An important contribution of our paper is that we do not restrict the functional form of the conditional selection probability. We provide semiparametric identification of selection-corrected quantile regression when the outcome is selectively observed and selection depends on unobservables. Identification is achieved by inverse probability weighting, where selection probabilities may depend on latent potential outcomes but are otherwise unaffected by the excluded variable. We establish identification of linear quantile regression even when nonparametric selection probabilities are not identified via conditional moment restrictions implied by the exclusion restriction. The constructive identification arguments motivate a flexible B-spline estimator, combined with cone projection, to estimate the initial selection probabilities. The resulting inverse probability quantile regression exhibits excellent finite-sample properties, and we establish its asymptotic behavior. 

We apply the proposed estimation strategy to derive selection-corrected wage distributions for men and women into full-time employment and to quantify the distribution of the selection-corrected gender wage gap among full-time workers. The empirical analysis is based on administrative social security data from Germany (SIAB), which provides detailed information on the complete employment and earnings histories of individuals in the labor force. We use the initial wage as the instrument, which captures persistent individual heterogeneity, such as ability, motivation or social skills, that affect potential earnings over the life cycle. This is in line with previous studies showing that initial wages have persistent effects on future outcomes (see, e.g., \cite{Deve_2002}, \cite{OreoWachter_2012}, \cite{SchwaWAcht_2019}). In the Roy model, the exclusion restriction implies that the instrument provides no additional information about an individual’s position in the reservation wage distribution. Intuitively, once potential wages and observed characteristics are held fixed, individuals with the same rank in the distribution of reservation wages remain comparable regardless of their value of the initial wage. 

Our results highlight the critical role of selection in shaping the gender wage gap. Women are positively selected into full-time employment across the wage distribution, with the strongest effects for lower and median wages. For men, selection effects are smaller but still positive and statistically significant. Conditional analyses yield a more nuanced picture. After controlling for education, experience, age, and workplace characteristics, selection correction consistently lowers women’s estimated wages, particularly at lower quantiles and among the less educated. This indicates that full-time working women in these groups tend to possess unobserved attributes (e.g., motivation, ability) associated with higher wages, leading to a positively selected sample. The selection effect diminishes at higher quantiles and is weakest among highly educated women, implying that top female earners are less affected by unobserved selection and that full-time workers in this group more closely represent the broader female population. For men, positive selection is also present across education groups but follows a different pattern. Among the highly educated, median wages decline by about 10.7 euros after correction, compared with 5.6 euros for women, resulting in a 21\% selection-corrected gender wage gap, which is 2 percentage points lower than the gender wage gap without selection correction. In contrast, selection effects are minimal among less educated men. These different selection patterns have important implications for the distribution of the gender wage gap. While the gender wage gap among the low educated increases from 3.6\% to 10.7\% at the median after correcting for selection, the selection-corrected gender wage gap for the high educated is slightly lower than the uncorrected wage gap. Overall, these results underscore the importance of accounting for selection heterogeneity across quantiles and education levels. Ignoring selection biases can mask substantial differences in the gender wage gap, particularly in the lower part of the wage distribution. 
In robustness checks, we impose stricter requirements on the lag structure of the instrument and find that the results remain virtually unchanged; see Appendix \ref{appx:robustness}.\\

\textit{Related Literature.} Our approach extends the existing literature on endogenous selection and nonignorable nonresponse, where instrumental variables are used to restore identification when selection depends on the outcome variable itself. IV-based strategies similar in spirit to ours have been developed in both the statistics (see, e.g., \cite{tang2003}, \cite{zhao2015}) and econometrics (see, e.g., \cite{2010Hault}, \cite{ramalho2013}, \cite{breunig2015}) literatures, as well as by \cite{breunig2020nonparametric} for the case of selectively missing covariates. More recent contributions by \cite{zhang_wang} and \cite{yu_etal} extend these ideas to quantile regression, proposing likelihood-based estimators under nonignorable selection that rely on parametric assumptions about the selection mechanism. In contrast, our method avoids such parametric restrictions, offering greater flexibility and robustness to model misspecification. While most applications of these statistical methods have focused on medical or clinical settings, we demonstrate their potential for economic contexts, particularly in analyzing labor supply and wage inequality.

By using an exclusion restriction for the underlying latent process, our strategy extends the existing literature analyzing the gender wage gap. Most previous studies focus on the average wage and address selection using the classical Heckman correction \citep{heckman1979sample}, which requires an exclusion restriction for the selection into employment. Typically, researchers have relied on variables such as the number and age of children (e.g., \cite{mulligan2008selection}) or on variation in the tax and transfer system, as suggested by \cite{BluZilLop_23}.\footnote{For a survey of studies using this approach, see \cite{blau2017gender} and \cite{Blau_etal_2021}.} However, these exclusion restrictions are often difficult to justify in practice, particularly when analyzing selection among men. Several studies have extended the classical Heckman model. \cite{das_newey_vella}, for example, develop a nonparametric approach and apply it to estimate returns to schooling among young Australian females. Of particular importance for our paper is the extension by \cite{arellano2017}, who propose a framework that allows for selection correction across the entire wage distribution, recognizing that selection patterns and their effects may vary across wage levels. Using UK data, they find positive selection for both men and women -- more pronounced for men -- and therefore report that the selection-corrected gender wage gap is smaller than the uncorrected one. Their identification strategy relies on the assumption that out-of-work benefit income is independent of unobserved determinants of wages. 

\cite{MaasWang_19} apply a similar approach for the US and show that selection-corrected gender wage gaps have increased over time, both at the mean and across various quantiles of the wage distribution. \cite{BluZilLop_23} focus on gender differences in the distribution of lifetime earnings. They find that gender disparities widened during the first half of individuals’ working lives but that substantial convergence occurred after age 40 for more recent cohorts, particularly compared to those born before the 1950s. Several alternative approaches have been proposed to study the distributional aspects of the gender wage gap (for an overview, see \cite{Blau_etal_2021}). 

\cite{blundell2007changes} develop and estimate bounds on the wage distribution to address non-random selection into employment. Although this method does not yield point identification, the resulting bounds are relatively tight. Their findings suggest a narrowing of the gender wage gap among the less educated. Relatedly, \cite{honore2020} examine identification in classical semiparametric sample selection models without exclusion restrictions. They derive sharp bounds for the parameters of interest and show that, even in the absence of valid instruments, the identified set can be narrow in practice.

\cite{Blau_etal_2021} further propose imputation-based methods to address missing wage information, following techniques similar to those in \cite{OliPet_08}. This approach has the advantage of correcting for selection across the full wage distribution. However, imputation methods rely solely on observed characteristics and are therefore based on the Missing at Random (MAR) assumption. Using such techniques, \cite{Blau_etal_2021} show that the reduction in the gender wage gap in the US over time becomes more pronounced when accounting for selection effects, although substantial disparities persist.\\
  
The paper is structured as follows. Section \ref{sec:model_identification} outlines the modeling framework and establishes semiparametric identification, introducing the main assumptions. We also illustrate the identification strategy using the Roy model and show under which conditions it is nested within our sample selection framework. Section \ref{sec:estimation} details the estimation procedure and derives the asymptotic properties of the estimator, as well as its finite-sample performance based on Monte Carlo simulations. The empirical application is discussed in Section \ref{sec:empirical_application}, including a description of the data, the empirical specification, and the instrument. Section \ref{sec:results} reports the empirical findings, and Section \ref{Sec:Conclusion} concludes. All proofs are provided in Appendices \ref{appx:proofs}--\ref{appx:cov_mat_est}. Appendix \ref{appx:outcome_var} describes the data preparation process and reports summary statistics for the wage variable at different stages of data cleaning. Appendix \ref{appx:cdf} illustrates the application of our estimator to the estimation of the distribution function. Appendix \ref{appx:simulation} contains additional simulation results and a detailed description of the simulation design.
Finally, Appendix \ref{appx:robustness} reports robustness checks.

\section{Semiparametric Identification}\label{sec:model_identification}

This section begins by outlining the assumptions required for identification. We then establish the identification of the quantile function using inverse selection probability weighting. We further extend our identification results more specifically to the Roy model of labor supply.
 
\subsection{Setup and Identification}\label{ssec:setup}

Given a latent outcome $Y^*$ (for example, potential earnings from employment) and the participation indicator $D$ (e.g., employment), we consider the following model for $\tau \in (0,1)$:
\begin{align}
    &Y^*= Z^\top \theta_\tau + U_\tau, \quad \text{where} \quad \PP(U_\tau \leq 0 \mid Z)= \tau\label{eq:main_model} \\
    &D=\mathbbm1\{V\leq p(Y^*,X)\}\label{selection:mod} \\
    &Y=Y^*\quad\text{if}\quad D=1,\nonumber
\end{align}
for some function $p(\cdot)$ that maps to the unit interval and where $\mathbbm1\{\cdot \}$ denotes the indicator function.
 Here, $U_\tau$ and $V$ are unobservable variables, and $Z= (1, X^\top,W^\top)^\top$, where $X$ is a vector of covariates and the random vector $W$ is excluded from the selection equation, which will be formally imposed below. We refer to $W$ as instruments. Throughout the paper, we assume that a random sample of $(Y,D,X^\top,W^\top)$ is available. In the following, we provide sufficient conditions to identify the quantile regression coefficient $\theta_\tau$, which is the $d_z$--dimensional parameter of interest. 

Identification is based on instrumental variables $W$ that explain variations in the latent variable $Y^*$ but are not directly related to the selection mechanism. This contrasts with parametric and semiparametric versions of the sample selection model of \cite{heckman1979sample}, which use variables that explain the selection equation but not the outcome (wage) equation, for example \cite{ahn1993}, \cite{Donald1995}, \cite{chen_khan}, \cite{das_newey_vella} and \cite{arellano2017}, the latter extending Heckman’s framework to quantile selection models.

\begin{assumption}
\label{A_assumptions_1-3}
(i) $V\independent W\mid (Y^*, X)$ and $V\mid (Y^*,X)\sim \mathcal U(0,1)$;
(ii) the selection probability \( p \) is uniformly bounded away from zero on its support; (iii) $p$ is identified through the conditional moment restriction \( \Ex[D / p(Y, X) \mid W,X] = 1 \). 
\end{assumption}

Assumption \hyperref[A1i]{\ref{A_assumptions_1-3}(i)} imposes conditional independence of the instrument $W$ and the unobservables $V$ in the selection equation. This assumption further imposes a uniform distribution of the conditional distribution of $V$, which is an innocuous normalization condition. Under Assumption \hyperref[A1i]{\ref{A_assumptions_1-3}(i)} we conclude that 
\begin{align*}
    \mathbb P(D=1\mid Y^*, X,W)=\mathbb P(V\leq p(Y^*, X)\mid Y^*, X)=p(Y^*,X).
\end{align*}
In particular, $p(Y^*,X)=\mathbb P(D=1\mid Y^*, X)$ by the law of iterated expectations, and thus we refer to $p$ as the selection probability in the remainder of the paper. 

Also note that Assumption \hyperref[A1i]{\ref{A_assumptions_1-3}(i)} excludes any relation between $W$ and the selection variable $D$ that is not channeled through $(Y^*,X)$. Assumption \hyperref[A1ii]{\ref{A_assumptions_1-3}(ii)} extends to the usual assumption of overlap required for the identification of population parameters under data missing at random, since herein the probability of selection can depend on the latent outcome. 

Throughout the paper, we define the inverse selection probability function as $g(y,x) := 1/\mathbb{P}(D = 1 \mid Y^* = y, X = x)$. Under Assumption \hyperref[A1ii]{\ref{A_assumptions_1-3}(i)(ii)}, this definition directly implies the following conditional moment restriction:
\begin{align}\label{cond:eq}
   \Ex[Dg(Y,X)\mid W,X]=1,
\end{align}
which, importantly, only involves observable random variables. Assumption \hyperref[A1iii]{\ref{A_assumptions_1-3}(iii)} ensures identifiability of the selection probability through this moment condition. In parametric settings, this requirement corresponds to the familiar rank condition, whereas in the nonparametric case it relies on a bounded completeness assumption, as shown by \cite{2010Hault}, which can be more challenging to justify. While our first result establishes identification of the quantile regression coefficient $\theta_\tau$ under Assumption \hyperref[A1iii]{\ref{A_assumptions_1-3}(iii)}, we show below that this assumption can also be avoided by imposing a mild variational condition on the quantile of interest.

Under the standard quantile regression assumption that, for a given $\tau \in (0,1)$, the conditional quantile function satisfies $Q_{Y^*}(\tau \mid Z = z) = z^\top \theta_\tau$ for some $\theta_\tau \in \mathbb{R}^{d_z}$, the parameter $\theta_\tau$ can be characterized, following the check function argument of \citet{KoenkerBassett}, as the solution to
\begin{equation}\label{eq:QR_equation}
    \theta_\tau = \argmin_{\theta \in \mathbb{R}^{d_z}}\Ex[\rho_\tau(Y^* - Z^\top \theta)],
\end{equation}
where $\rho_\tau(u) = u(\tau - \1\{u < 0\})$ denotes the check function. Under Assumption~\ref{A_assumptions_1-3}, this parameter is identified and uniquely determined, as stated in the following proposition.
\begin{prop}\label{lem:identification}
Suppose Assumption \ref{A_assumptions_1-3} holds. Then for any $\tau\in(0,1)$, the quantile regression coefficient $\theta_\tau$ is uniquely determined as the solution to
\begin{equation}\label{eq:theta_tau}
    \theta_\tau = \argmin_{\theta \in \mathbb{R}^{d_z}} \, \mathbb{E}\left[ D \, g(Y,X) \, \rho_\tau\big(Y - Z^\top \theta\big) \right].
\end{equation}
\end{prop}

The next theorem establishes identification of our quantile selection model even if the inverse selection probability function $g$ is not identified through the conditional moment condition \eqref{cond:eq}, as imposed in Assumption \hyperref[A1iii]{\ref{A_assumptions_1-3}(iii)}. For this result, we adopt an insight from \cite{severini_tripathi2012} regarding point identification of functionals of partially identified structural functions. 
\begin{theorem}\label{prop:1}
Suppose Assumption \ref{A_assumptions_1-3}(i)(ii) holds. Then, for any $\tau \in (0,1)$ for which a square integrable function $\mu_\tau$ exists that satisfies
\begin{equation}\label{eq:primitive}
    \Ex\left[\mu_\tau(W,X)\mid Y^*,X\right]=\Ex\left[\rho_\tau\big(Y^* - Z^{\top} \theta_\tau\big) \bigm| Y^*,X\right],
\end{equation}
the quantile regression coefficient $\theta_\tau$ is identified by \eqref{eq:theta_tau}. 
\end{theorem}
Equation \eqref{eq:primitive} requires that the instrument $W$ carry sufficient information so that the conditional expectation $\Ex\left[\rho_\tau\left(Y^* - Z^{\top} \theta_\tau\right) \mid Y^*,X\right]$ can be expressed as an expectation of some measurable function of $(W,X)$ given $(Y^*,X)$. In other words, the assumption postulates the existence of a bridge function linking the residual loss function to the observed variables. This corresponds to a proximal identification requirement, where, conditional on $X$,  the instrument $W$ spans the same information space for the relevant conditional moment as the latent outcome $Y^*$ does.

\subsection{Application to the Roy Model}\label{ssec:roy_model}
To illustrate identification, we use the Roy model \citep{Roy_51} applied to the decision on labor supply by \cite{gronau1974} and \cite{heckman1974}.\footnote{\cite{mulligan2008selection} define this model as the Gronau-Heckman-Roy (GHR) labor supply model.}
In this model, $Y^*$ is the potential wage of employment and $D$ is the indicator of labor market participation. 
According to the labor supply model, individuals choose employment if their potential wage $Y^*$ exceeds their reservation wage $R$: 
\begin{align}\label{roy:model}
    D=\mathbbm1\{Y^* \geq R\}.
\end{align}
In the following, we explicitly derive under which conditions the Roy model is nested within our sample selection framework.
 
We assume that the reservation wage $R$ is continuously distributed conditional on the potential wage $Y^*$ and the observed characteristics $X$. Formally, the cumulative distribution function $F_{R\mid Y^*,X}(\cdot)$ is continuous and strictly increasing.

In the Roy model, labor market participation is determined by the comparison of potential and reservation wages:
\[
\mathbb P(D=1\mid Y^*,X)=\mathbb P(Y^*\geq R \mid Y^*,X)=F_{R\mid Y^*,X}(Y^*).
\]
Since $F_{R\mid Y^*,X}(\cdot)$ is strictly increasing, we can interpret $F_{R\mid Y^*,X}(R)$ as the \emph{conditional rank} (or quantile position) of the reservation wage within the distribution of reservation wages among individuals with the same $(Y^*,X)$. Under the continuity assumption, this rank is uniformly distributed on $(0,1)$ conditional on $(Y^*,X)$.

Our key independence condition is that this reservation-wage rank is independent of the instrument $W$ once we condition on the potential wage and observed characteristics:
\begin{align}\label{eq:rank_invariance}
    F_{R\mid Y^*,X}(R) \;\independent\; W \mid (Y^*,X).
\end{align}
This assumption states that, conditional on productive ability and observable characteristics, the instrument does not provide information about unobserved determinants of labor supply decisions. It thus rules out selection on unobservables with respect to the instrument while allowing for rich and continuous heterogeneity in reservation wages.
This condition can be interpreted as a \emph{rank invariance} assumption (on the reservation wage), analogous to that used in quantile IV models (see \cite{Chern2005}). It relaxes the conditional independence assumption on the reservation wage itself, imposed in Example 2 in \cite{2010Hault}.  The rank invariance condition \eqref{eq:rank_invariance}  implies that, given $(Y^*,X)$, the instrument $W$ does not provide additional information about an individual’s position in the reservation wage distribution. Intuitively, once the potential wages and observed characteristics are fixed, individuals with the same rank in the distribution of reservation wages remain comparable regardless of their value of $W$.
Consequently, we obtain the following identification result.
\begin{coro}\label{coro:roy_model}
    For the Roy model \eqref{roy:model}, suppose that Assumption \ref{A_assumptions_1-3}(ii) holds and that $R$ given $(Y^*,X)$ is continuously distributed. In addition, assume that the rank invariance condition \eqref{eq:rank_invariance} is satisfied. Then, for any $\tau \in (0,1)$ that satisfies \eqref{eq:primitive} the quantile regression coefficient $\theta_\tau$ is identified by \eqref{eq:theta_tau}. 
\end{coro}

In our empirical application (see Section \ref{sec:empirical_application}), we use an individual's initial wage as the instrument $W$. The initial wage captures unobserved individual heterogeneity, such as ability, diligence, and both hard (education, technical expertise) and soft (teamwork, communication, negotiation) skills, that shape potential wages over the life cycle. This argument is consistent with explanations in the literature showing that initial wages have persistent effects on future outcomes (see, e.g., \cite{Deve_2002}, \cite{OreoWachter_2012}, \cite{SchwaWAcht_2019}). For example, \cite{OreoWachter_2012} show that career development models or models of human capital accumulation can explain why high initial wages lead to persistent higher wages.

The exclusion restriction in Corollary \ref{coro:roy_model} implies that, once we fix $(Y^*,X)$, the initial wage $W$ does not carry  further information on an individual’s position in the conditional distribution of reservation wages. In other words, among individuals with the same potential wage and observable characteristics, the initial wage does not predict whether someone is at a high or low quantile of the reservation wage distribution. Note that this exclusion restriction is far less restrictive than assuming that the initial wage has no effect on the reservation wage. In fact, the exclusion restriction is not violated if the initial wage would shift the level of the conditional reservation wage while not changing the rank distribution.

In the empirical application, we discuss in detail how we construct the initial wage and argue that it is important to use wage information of the distant past to relax strong assumptions of a static labor supply model. For details, see Section \ref{ssec:instrument}.

\section{Estimation and Inference} \label{sec:estimation}
In this section, we sketch an estimation procedure that builds on the identification result. We then analyze the asymptotic properties of the estimator and evaluate its finite-sample performance through simulation evidence.
\subsection{Estimation} \label{ssec:estimation}
Following the identification result, the estimation of the conditional quantiles of the latent outcome variable $Y^*$ proceeds in three steps, summarized in Algorithm~\ref{alg:est}.

The first step estimates the inverse selection probability function $g(Y, X)$ using the moment condition:
\[
\mathbb{E}\left[D g(Y, X) \mid W,X \right] = 1, \quad \text{where } Y = D Y^*.
\]

Let $\phi^J(\cdot)$ and $b^K(\cdot)$ be vectors of basis functions of dimensions $J$ and $K \geq J$, respectively. Denote by $\Phi_J$ the linear span of $\{\phi_1, \dots, \phi_J\}$. We assume that the function $g(Y,X)$ lies in this linear span, i.e., $g(Y, X) = \phi^J(Y, X)^\top \beta$ for some coefficient vector $\beta \in \mathbb{R}^J$, where $\phi^J(Y, X)$ denotes a transformation of $(Y,X)$. Multiplying both sides of the moment condition by the spline basis functions of $(W,X)$ yields:
\begin{equation} \label{eq:2SLS}
    \Ex\left[Db^K(W,X)\phi^J(Y,X)^\top\right]\beta = \Ex\left[b^K(W,X)\right].
\end{equation}

In the second step, we enforce a lower bound on the estimated inverse selection probabilities by projecting the unconstrained estimator $\widehat{g}_u$ onto the cone of functions bounded below by 1. This ensures $\widehat{g}(y,x)\geq1$ for all $(y,x)$. The projection is implemented using the \texttt{coneproj} package in R. Cone projection of series TSLS estimators for constrained hypothesis testing was also considered by \cite{breunig_chen_2024}. 

In the last step, we estimate the conditional quantile function of the latent outcome variable $Y^*$ via weighted quantile regression, using weights $D\,\widehat{g}(Y,X)$. This extends the quantile regression framework of \citet{KoenkerBassett} to account for selection, with inverse probability weights that depend jointly on the selection indicator and the outcome. For the algorithm, let $\mathcal{Y} \times \mathcal{X}$ denote the support of $(Y,X)$, and let $\mathbb{E}_n[\cdot]$ denote the empirical mean. For $(Y,X)$, we define $L^2(Y,X)=\{h: \left\lVert h\right\lVert_{L^2(Y,X)}<\infty\}$, where $\left\lVert h\right\lVert_{L^2(Y,X)} := \sqrt{\Ex[h^2(Y,X)]}$. The conditional probability density function of a random variable $Y$ given $Z$ is denoted by $f_{Y|Z}(\cdot)$. Throughout the paper, we assume that a sample $\{S_i\}_{i=1}^n$ is observed, where $S_i = (D_i, Y_i, W_i^\top, X_i^\top)$.

\FloatBarrier
\begin{algorithm}
\caption{Estimation Procedure with Inverse Probability Weighting}\label{alg:est}
\begin{algorithmic}[1]
\vspace{0.5em}
\Statex \textbf{Input:} Sample data $\{S_i\}_{i=1}^n$ and quantile level $\tau \in (0,1)$.
\vspace{0.5em}

\State \textbf{Estimate the inverse selection probability function.}  
Obtain the unconstrained estimator $\widehat{g}_u(\cdot) = \phi^J(\cdot)^\top \widehat{\beta},$
where $\widehat{\beta}$ is the 2SLS coefficient regressing the constant 1 on $D\phi^J(Y,X)$ with $b^K(W,X)$ as instruments.
\vskip0.6em
\State \textbf{Impose the shape restriction.}  
Project $\widehat{g}_u$ onto the cone $\mathcal{C} := \{g \in L^2(Y,X): g(y,x) \geq 1 \text{ for all } (y,x)\in\mathcal Y\times\mathcal X\}$ by solving
\[
\widehat{g} = \argmin_{h \in \mathcal{C} \cap \Phi_J} \, \mathbb{E}_n \Big[ \bigl(\widehat{g}_u(Y, X) - h(Y,X)\bigr)^2 \Big].
\]

\State \textbf{Estimate the conditional quantile function of $Y^*$.}  
Estimate the quantile parameter $\theta_\tau$ by
\begin{equation} \label{eq:theta_hat}
    \widehat{\theta}_\tau = \argmin_{\theta \in \mathbb{R}^{d_z}}\mathbb{E}_n \Big[ D \, \widehat{g}(Y,X) \, \rho_\tau(Y - Z^\top \theta) \Big].
\end{equation}

\Statex \textbf{Output:}  
Estimates $\widehat{\theta}_\tau$ and conditional quantile function $\widehat{Q}_{Y^*}(\tau \mid Z=z)= z^\top \widehat{\theta}_\tau$.
\end{algorithmic}
\end{algorithm}

\subsection{Asymptotic Properties} \label{ssec:asymptotics}
In this subsection, we present the asymptotic distribution for the quantile selection estimator $\widehat{\theta}_{\tau}$. To establish the result, we introduce some notation. Let $\Omega_g := D g(Y, X)$ be the selection weights and let $\psi_\tau(u) := \tau - \1\{u < 0\}$ be the quantile score function. Define $T:= \mathbb{E} \left[ \phi^J(Y,X) D Z^{\top} \psi_\tau(U_\tau)\right],$ where $U_\tau = Y^*-Z^\top \theta_\tau$.\footnote{Note that we can replace $Y^*$ by $Y$ whenever $U_\tau$ is multiplied by $D$, as $Y = D Y^*$} The matrices $G$ and $H$ capture key population moments involving the basis functions and the first-stage estimation errors: $G := \Ex[b^K(W,X)b^K(W,X)^\top ]$, $H := \Ex[D\phi^J(Y,X) b^K(W,X)^\top]$. Let $U_J := 1 - D\phi^J\left(Y,X\right)^\top\beta$ be the first-stage residual.
To obtain our asymptotic distribution result, we assume that the coefficient $\beta$ is identified through the moment condition \eqref{eq:2SLS}. 
Under additional assumptions as specified in Appendix \ref{appx:thm3proof}, the influence function, as derived in the next theorem, can be written as: 
\begin{equation*}
    \chi_{\tau g}(S_i):=M_{1\tau g}^{-1}\left( Z_i\Omega_{gi} \psi_\tau\left(U_{\tau i}\right)+T^\top(H G ^{-1} H^\top)^{-1} H G ^{-1}b^K(W_i,X_i)U_{Ji}\right),
\end{equation*}
where $M_{1\tau g} := \Ex \big[ \Omega_g f_{Y\mid\Omega_g, Z}\left(Z^\top \theta_\tau\right)Z Z^{\top} \big].$
We now present the asymptotic distribution of our quantile selection estimator $\widehat{\theta}_\tau$.

\begin{theorem}
\label{thm:asymptotic_normality}
Suppose Assumptions \ref{A_assumptions_1-3}(i)(ii), \ref{A_asymptotics2}, and \ref{A_asymptotics1} hold. Then, for any $\tau \in (0,1)$ that satisfies \eqref{eq:primitive} we have
$$
\sqrt{n}\left(\widehat{\theta}_{\tau}-\theta_{\tau}\right) = \frac{1}{\sqrt{n}}\sum_{i=1}^n \chi_{\tau g^*}(S_i) + o_p(1)
$$
and in particular
$$
\sqrt{n}\left(\widehat{\theta}_{\tau}-\theta_{\tau}\right) \xrightarrow{d} \mathcal{N}\left(0, \Ex\big[\chi_{\tau g^*}(S)\chi_{\tau g^*}^\top(S)\big]\right),
$$
where $g^*$ satisfies \eqref{cond:eq} and minimizes $\Ex\big[\chi_{\tau g}(S)\chi_{\tau g}^\top(S)\big]$.
\end{theorem}

This result provides the asymptotic distribution of the quantile regression estimator. The detailed regularity conditions required to establish this asymptotic result, together with its proof, are given in Appendix~\ref{appx:thm3proof}. The asymptotic covariance estimator used in our implementation is presented in Appendix~\ref{appx:cov_mat_est}.\footnote{In our application, we estimate the conditional densities using the \texttt{npcdens} package in R, which implements the method of \cite{li_racine_2004}, with bandwidths selected via the cross-validation procedure of \cite{hall_li_racine_2004}.}

\subsection{Monte Carlo Studies}\label{sec:simulation}

We evaluate the finite-sample properties of the estimator compared to existing approaches, using 1{,}000 replications with $n=1{,}000$ observations per replication.

We consider the outcome model
\[Y^*_i = \beta_0 + \beta_1 W_{i} + \beta_2 X_{i} + \epsilon_i(\tau),\quad i = 1, \ldots, n,\]
where $(W_{i}, X_{i})^\top \sim \mathcal{N}\left(\begin{pmatrix}
2 \\ 1 \end{pmatrix},\begin{pmatrix} 1 & 0.5\\ 0.5 & 1 \end{pmatrix}\right)$ and $(\beta_0, \beta_1, \beta_2) = (1,1,2)$. The error term $\epsilon_i(\tau)$ ensures a zero conditional $\tau$-quantile. In this section, we present results for $\epsilon_i(\tau)$ drawn from a t-distribution with 3 degrees of freedom, scaled by 0.7, while sensitivity to alternative error distributions is presented in Appendix~\ref{appx:simulation}.

Selection into the observed sample is modeled through a binary indicator $D_i$, which equals one if individual $i$ is observed and zero otherwise. Conditional on the latent outcome $Y_i^*$ and covariates $X_i$, we assume a logistic selection model:
\begin{equation*}
    D_i \sim \text{Bernoulli}(p(Y_i^*, X_i)), \quad \text{where } p(Y_i^*,X_i) = \frac{1}{1 + \exp(-s(Y^*_i,X_i))}.
\end{equation*}
The function $s(\cdot)$ determines the selection mechanism. We consider three specifications that differ in how selection depends on observed covariates and the latent outcome. In the first mechanism (M1, MAR), selection depends only on observed covariates, with $s(Y^*_i, X_i) = \alpha_1 + \gamma_1 X_{i}$. The second (M2, MNAR-linear) introduces dependence on the latent outcome in a linear form, $s(Y_i^*, X_i) = \alpha_2 + \gamma_2 X_{i} + \xi_2 Y_i^*$. The third mechanism (M3, MNAR-nonlinear) allows for nonlinear dependence on the covariate, $s(Y_i^*, X_i) = \alpha_3 + \gamma_3 \sin^2(X_{i}) + \xi_3 Y_i^*$. The parameters $(\alpha_j, \gamma_j, \xi_j)$ are chosen such that the average missing-data rate is approximately 35\%. The design mirrors our empirical application: $W$ serves as an instrumental variable that affects the latent outcome but is excluded from the selection equation, paralleling the role of initial wages as instruments in our application. The exact parameter values and further implementation details are provided in Appendix~\ref{appx:simulation}.

We compare four estimation methods at the median quantile, $\tau = 0.5$. First, the uncorrected complete-case estimator relies solely on the observed outcomes, ignoring the missing-data mechanism and therefore serving as a naive benchmark. Second, the MAR correction assumes that selection depends only on observable characteristics and is implemented through a two-step inverse probability weighting procedure. Third, we include the joint estimating equations (JEE) method of \citet{yu_etal}, which, similar in spirit to ours, employs an IV but specifies a parametric form for the selection probability and jointly estimates the quantile and selection equations via an augmented likelihood framework. Finally, our semiparametric IV method corrects for selection using an instrument without imposing any parametric restrictions on the selection probability, providing greater flexibility and robustness to model misspecification.

Our estimation procedure follows Algorithm~\ref{alg:est}. In the first step, we estimate $\widehat{g}_u(Y,X)$ using quadratic B-splines -- without interior knots for $Y$ and with two interior knots for $X$ -- yielding three and five basis functions, respectively. Including one control variable, this yields dimensions $J=4$ and $K=6$. In the second step, we impose the lower-bound shape constraint using the \texttt{coneproj} package in R to obtain the constrained estimate $\widehat{g}(Y,X)$. Finally, quantile coefficients are estimated using weighted quantile regression with weights $D\widehat{g}(Y,X)$. The covariance matrix and confidence intervals are computed using the plug-in estimators detailed in Appendix~\ref{appx:cov_mat_est}.

\FloatBarrier
\begin{table}[ht]
\centering 
    \caption{(Setting C) Simulation results by method with $(\tau, n) = (0.5, 1000)$ and 35\% missing in the outcome}
    \label{tab:settingCresults}
    \resizebox{0.99\linewidth}{!}{%
\begin{tabular}[t]{lrrr|rrr|rrr}
\toprule
\multirow{2}{*}{Method} & \multicolumn{3}{c|}{M1} & \multicolumn{3}{c|}{M2} & \multicolumn{3}{c}{M3}\\
& \multicolumn{1}{c}{$\beta_0$} & \multicolumn{1}{c}{$\beta_1$} & \multicolumn{1}{c|}{$\beta_2$} & \multicolumn{1}{c}{$\beta_0$} & \multicolumn{1}{c}{$\beta_1$} & \multicolumn{1}{c|}{$\beta_2$} & \multicolumn{1}{c}{$\beta_0$} & \multicolumn{1}{c}{$\beta_1$} & \multicolumn{1}{c}{$\beta_2$} \\
\midrule
& \multicolumn{9}{c}{Mean biases} \\
\cmidrule{2-10}
Uncorrected & 0.005 & -0.002 & 0.001 & 0.249 & -0.025 & -0.076 & 0.255 & -0.025 & -0.077 \\
MAR-assumed & 0.004 & -0.001 & 0.001 & 0.317 & -0.033 & -0.108 & 0.296 & -0.034 & -0.089 \\
JEE & 0.001 & -0.004 & 0.005 & 0.010 & -0.012 & 0.000 & 0.005 & 0.001 & -0.008 \\
Semiparametric IV & -0.033 & 0.009 & 0.012 & 0.042 & 0.005 & -0.018 & 0.056 & 0.003 & -0.029\\
\addlinespace
\cmidrule{2-10}
& \multicolumn{9}{c}{RMSE} \\
\cmidrule{2-10}
Uncorrected& 0.094 & 0.044 & 0.043 & 0.271 & 0.048 & 0.090 & 0.275 & 0.050 & 0.090\\ 
MAR-assumed & 0.100 & 0.047 & 0.047 & 0.377 & 0.072 & 0.156 & 0.362 & 0.077 & 0.135\\ 
JEE & 0.120 & 0.102 & 0.125 & 0.841 & 0.362 & 0.330 & 0.714 & 0.265 & 0.179\\
Semiparametric IV & 0.110 & 0.047 & 0.048 & 0.131 & 0.050 & 0.060 & 0.133 & 0.049 & 0.064\\
\addlinespace
\cmidrule{2-10}
& \multicolumn{9}{c}{CI lengths} \\
\cmidrule{2-10}
Uncorrected & 0.355 & 0.170 & 0.178 & 0.408 & 0.170 & 0.186 & 0.402 & 0.171 & 0.184 \\ 
MAR-assumed & 0.371 & 0.175 & 0.191 & 0.668 & 0.230 & 0.382 & 0.663 & 0.250 & 0.357 \\ 
JEE & 0.716 & 0.306 & 0.332 & 1.063 & 0.406 & 0.495 & 0.994 & 0.360 & 0.447 \\ 
Semiparametric IV & 0.701 & 0.344 & 0.318 & 0.636 & 0.316 & 0.256 & 0.674 & 0.327 & 0.263\\
\addlinespace
\cmidrule{2-10}
& \multicolumn{9}{c}{Coverage probabilities} \\
\cmidrule{2-10}
Uncorrected & 0.943 & 0.944 & 0.963 & 0.343 & 0.921 & 0.639 & 0.302 & 0.903 & 0.626\\
MAR-assumed & 0.935 & 0.932 & 0.961 & 0.409 & 0.891 & 0.657 & 0.480 & 0.885 & 0.709 \\
JEE & 0.941 & 0.935 & 0.955 & 0.879 & 0.927 & 0.882 & 0.854 & 0.924 & 0.882\\ 
Semiparametric IV & 0.989 & 1.000 & 0.990 & 0.965 & 0.994 & 0.956 & 0.977 & 0.993 & 0.948\\
\bottomrule
\bottomrule
  \label{tab:settingC}

\end{tabular}}
\end{table}

While the full set of simulation results and the detailed description of the setup -- covering all error distributions and selection mechanisms -- are provided in Appendix~\ref{appx:simulation}, Table \ref{tab:settingC} summarizes results for a representative setting with heavy-tailed errors: $t$-distributed errors with 3 degrees of freedom, scaled by 0.7. The proposed semiparametric IV estimator exhibits negligible bias and low RMSE across all selection mechanisms. Under M1, where selection is independent of the latent outcome given covariates, all estimators perform comparably. The proposed method tends to produce slightly wider confidence intervals, reflecting its robustness against more complex forms of selection. In contrast, when selection depends on the latent outcome (M2 and M3), both the MAR-assumed and complete-case estimates suffer from substantial bias, high RMSE, and poor coverage. The joint EE method also fails under M2 and M3, with sharp increases in RMSE and confidence interval length. Overall, the semiparametric IV method demonstrates the most robust performance, consistently achieving the lowest RMSE and the highest coverage even in the presence of selection dependent on the latent outcome. Moreover, the joint EE method is computationally more intensive, as it relies on bootstrapping to estimate standard deviations. For large samples, as in our empirical application, this makes it difficult to implement the estimator, whereas our method is considerably faster.

\section{Empirical Application}\label{sec:empirical_application}

In this section, we present the empirical application and outline the implementation procedure. We begin by describing the data and providing summary statistics of the key variables. Next, we discuss the construction and choice of the instrumental variable. We then detail the implementation of our proposed selection correction method, explaining how it is applied to the data alongside the benchmark estimations -- both the uncorrected model and the selection correction under the MAR assumption.

\subsection{Data Description and Summary Statistics}\label{ssec:data_description}

The empirical analysis is based on administrative social security data from Germany.\footnote{We use the factually anonymous version of the Regional File of the Sample of Integrated Labor market Biographies 1975-2017 (SIAB-R 7517), a 2\% random sample drawn from the Integrated Employment Biographies (IEB), provided by the Institute for Employment Research (IAB) \citep{antoni2019sample}.} The data include individuals who have held employment subject to social security contributions or marginal part-time employment at least once during the observation period, as well as benefit recipients and job seekers.\footnote{Marginal part-time employment is recorded from 1999 onward. Following the German unification, for East Germany recorded information is assumed to be complete from 1993 onward.} It provides comprehensive information on the employment history, including employment spells, daily wages, and various demographic characteristics such as gender, age, education, work experience, and occupation. 

Our analysis is based on the 2017 wave of the dataset. We focus on individuals between the ages of 25 and 50 who have held German nationality throughout their entire recorded labor history. Our sample includes i) individuals in part-time and full-time employment, ii) recipients of unemployment benefits and means-tested transfers, and iii) job seekers not receiving unemployment benefits. We exclude individuals who are still in education or vocational training, and those who are partially retired. Additionally, our analysis excludes self-employed individuals and civil servants, as the administrative records do not cover these employment groups. Finally, we do not observe individuals who are part of the working-age population but are not in the labor force because they are not actively seeking employment. Thus, we analyze the effects of non-random selection of individuals in the labor force into full-time employment. 

The administrative records offer three major advantages for analyzing gender-specific wage distributions, and the distribution of the gender wage gap for full-time workers. First, the data cover a broad and representative segment of the population with a large sample size, enabling the use of nonparametric estimation techniques, which typically require larger datasets than parametric methods.\footnote{While the large sample size is beneficial, our estimation strategy remains applicable even in smaller samples.} Second, the data are extensive and precise. Unlike self-reported survey data, these records avoid issues of misreporting and non-random missing observations in wage data. Finally, the data include information about the full earnings and working history with occupational details, which is required to construct the instrument.

Our key variable of interest is the gross daily wage in full-time employment, which is recorded for every spell of full-time employment.\footnote{The data do include information about working hours. Therefore, we focus on the gender wage gap in full-time employment.} To construct a panel by year and individual, we closely follow \citet{dauth2020preparing}: we record individuals with spells spanning June 30th of each year in the panel and use the wage from the longest employment spell within the respective year as the gross daily earnings for that year. The gross daily wage is top-coded due to social security insurance limits, meaning wages above this threshold are not fully observed.\footnote{The limit's assessment ceiling varies by year and between East and West Germany. In 2017, it was 76,200 euros in West Germany and 68,400 euros in East Germany.} This censoring restricts the analysis of the full wage distribution. Therefore, we trim the top 5\% of the wage distribution by gender, year, and region, removing individuals whose earnings ever fall within the top 5\%. Additionally, to avoid outliers, we trim the bottom 5\% of the wage distribution. The data preparation process and the sample restriction are discussed in detail in Appendix \ref{appx:outcome_var}. Overall, we use 119,930 observations for females and 124,588 observations for males in our empirical analysis.

Table \ref{tab:descriptive_stats} provides information about gross daily wages in full-time employment and an overview of the sample's descriptive statistics.\footnote{Wages are adjusted for inflation, using 2015 as the base year.} Consistent with previous evidence, we observe a clear raw gender wage gap in full-time earnings. On average, men earn about 30 euros more per day than women, corresponding to a raw gender wage gap of approximately 25\%. The median wage gap is slightly smaller, around 20\%.\footnote{The gender gap in full-time earnings in our data is slightly higher than that reported in \cite{IlievaWrohlich_22}, who focus on hourly wages based on SOEP data.} 

Education is classified into three groups based on the completion of a university degree or vocational training.\footnote{For more details, see \citet{dauth2020preparing}. Note that the proportion of university graduates in the general population is higher than in our sample, as the top 5\% of the wage distribution has been trimmed, potentially excluding individuals with higher educational qualifications.} Educational attainment is generally similar across genders: the majority have completed vocational training, while a smaller proportion hold a university degree.

\begin{table}[h!]\centering\footnotesize
    \begin{threeparttable}
    \caption{Descriptive Statistics} \label{tab:descriptive_stats}
    \begin{tabular}{lcccc}
    \toprule
                        &\multicolumn{2}{c}{Women}             &\multicolumn{2}{c}{Men}               \\\cmidrule(lr){2-3} \cmidrule(lr){4-5}
                    &\multicolumn{1}{c}{Mean/Share} & \multicolumn{1}{c}{Median} &\multicolumn{1}{c}{Mean/Share} & \multicolumn{1}{c}{Median}\\
\midrule

Gross Daily Full-Time Wage (in Euro)   &    91.26 & 91.18 & 112.86 & 107.84 \\ \vspace{-0.7em}
               &      &    &    &  \\
Employment Status  (\%) &        &    &       &      \\
\hspace{2mm} Non-employed&        6.23&  & 6.67&\\
\hspace{2mm} Part-time&       52.18 & \multicolumn{1}{c}{\checkmark}& 10.09 &\\
\hspace{2mm} Full-time &       41.59& & 83.24 & \multicolumn{1}{c}{\checkmark} \\
Age                 &       38.74 &   39.00 &    37.95 & 38.00 \\
Education Level (\%)           &    &        &      & \\
\hspace{2mm} No vocational training &        5.18&  &   5.55 \\
\hspace{2mm} Vocational training &       77.88&  \multicolumn{1}{c}{\checkmark}  &   77.79 & \multicolumn{1}{c}{\checkmark}\\
\hspace{2mm} University&       16.94&  &     16.66& \\
Workplace (\%)           &    &        &      & \\
\hspace{2mm} Non-employed &   6.23&  &   6.67 \\
\hspace{2mm} Employed in East Germany &  17.70 &   &   17.35 & \\
\hspace{2mm} Employed in West Germany &     76.07 & \multicolumn{1}{c}{\checkmark} &     75.97& \multicolumn{1}{c}{\checkmark}\\
Experience (in years)         &   &         &      &      \\
\hspace{2mm} Non-employed&        0.77 & 0.00 & 0.87 & 0.00\\
\hspace{2mm} Part-time&        4.86&  3.00& 0.96 & 0.00 \\
\hspace{2mm} Full-time&        7.30&  6.00& 11.59 & 10.00\\
Earliest job difficulty   (\%)       &   &         &      &      \\
\hspace{2mm} Unskilled/semiskilled task&     6.38 &  & 6.75 & \\
\hspace{2mm} Skilled task &        80.03 & \multicolumn{1}{c}{\checkmark} & 82.11 & \multicolumn{1}{c}{\checkmark}\\
\hspace{2mm} Complex task&        4.22 &  & 3.65 & \\
\hspace{2mm} Highly complex task &    9.37 &  & 7.49 & \\

\midrule
Individuals         &\multicolumn{2}{c}{119,930}&  \multicolumn{2}{c}{124,588} \\
    \bottomrule
    \end{tabular}
    \begin{tablenotes}[flushleft] \footnotesize
    \item \textit{Notes: SIAB-R data, 2017 cross-section. The sample includes individuals aged 25-50, all German nationals. Education reflects the highest degree attained. Employment status is based on employers' mandatory social security notifications. Wages represent gross daily full-time earnings from each individual’s longest employment spell in 2017.}
\end{tablenotes}
    \end{threeparttable}
\end{table}

The employment status is derived from mandatory social security notifications submitted by employers. The distinction between full- and part-time work is based on a comparison of contracted hours with the standard working hours at the respective establishment. Non-employment spells are identified through the receipt of unemployment benefits or means-tested transfers. A notable disparity in employment status is observed: more than half of women (52\%) work part-time, compared to only 10\% of men. In contrast, the majority men (83\%) are employed full-time, while only 42\% of women are. The share of individuals in the labor force but currently out of employment is similar for men and women. To account for regional differences in labor market trajectories, we classify individuals by workplace region, distinguishing between East and West Germany, which evolved differently after reunification. In 2017, around 76\% of both men and women in our sample were employed in West Germany.

The data enable us to reconstruct detailed employment and wage histories for all individuals. We aggregate past employment into part-time and full-time experience and account for periods of non-employment.\footnote{Some unrecorded periods or data gaps may reflect time spent abroad or other forms of employment. Since we cannot verify these activities, we do not classify them as full-time, part-time, or non-employment.} Table \ref{tab:descriptive_stats} shows that women have significantly more part-time work experience -- five years on average compared to one year for men -- while men have more full-time experience, averaging 11.6 years versus 7.3 years for women. The median part-time experience for men is zero, whereas women have accumulated a median of three years.\footnote{In the empirical analysis, we control for experience but do not interpret the coefficients. Experience is highly correlated with age, and part-time employment is often linked to the presence and age of children, which are not observed in the data, making a meaningful interpretation challenging.} To account for initial career starting points, occupational differences, and potential variation in wage trajectories across occupations, we classify individuals based on the difficulty level of their earliest job. This job also determines their initial wage, which we use as an instrument in our analysis.

\subsection{Instrument}\label{ssec:instrument}

As mentioned above, we use an instrument $W$ based on the individual's wage history constructed to satisfy the exclusion restriction. The instrument explains variation in the potential wage but, conditional on this latent outcome and observable characteristics, does not directly affect the employment decision.
We argue that early wage history satisfies this restriction because it captures persistent individual heterogeneity -- such as ability, diligence, and both hard (education, technical knowledge) and soft (teamwork, communication, negotiation) skills -- that shape potential wages over the life cycle. Under this assumption, the initial wage influences labor market participation only indirectly, through its effect on the potential wage. This interpretation aligns with theories of career development, which suggest that higher initial wages provide greater opportunities for human capital accumulation and are therefore positively correlated with future skill acquisition and long-term earning potential \citep{OreoWachter_2012}.

\FloatBarrier
\begin{table}[h!]\centering\footnotesize
  \begin{threeparttable}
    \caption{Instrument: Initial wage}\label{tab:descript_instr}
    \begin{tabular}{lcccccc}
    \toprule
    &\multicolumn{3}{c}{Women}             &\multicolumn{3}{c}{Men}               \\\cmidrule(lr){2-4}\cmidrule(lr){5-7}
&\multicolumn{3}{c}{}                  &\multicolumn{3}{c}{}                  \\[-1em]
&\multicolumn{1}{c}{Mean}&\multicolumn{1}{c}{Min}&\multicolumn{1}{c}{Max}&\multicolumn{1}{c}{Mean}&\multicolumn{1}{c}{Min}&\multicolumn{1}{c}{Max}\\
\midrule
&            &            &            &            &            &            \\
\hspace{2mm} Initial wage (in euro) &        50.26&      4.29&        150.00&      60.25&       4.46&       192.81\\
\hspace{2mm} Age at initial wage    &       22.22&       17&       48&       21.97&       17&       48\\
\hspace{2mm} Years to 2017&       16.52&        2&       33&       15.97&        2&       33.00\\
\hspace{2mm} Share full-time (\%)&       93.26&        &  &       98.38&        &      \\
\midrule
Individuals         &\multicolumn{1}{c}{119,930}&            &            &\multicolumn{1}{c}{124,588}&            &            \\
    \bottomrule
    \end{tabular}
   \begin{tablenotes}[flushleft] \footnotesize
	\item \textit{Notes: SIAB-R data, 1975-2017. The initial wage is the daily wage at the earliest available point in the individual’s wage history within the SIAB dataset.}
    \end{tablenotes}
  \end{threeparttable}
\end{table}

Arguably, the assumptions of a static labor supply model may be violated due to factors such as state dependence in preferences \citep{heckman1991}, labor market dynamics, wage persistence \citep{MegPis_11}, or intertemporal financial incentives. Specifically, wages from recent periods may influence current employment decisions. For example, unemployment insurance payments or other transfers are often determined by recent earnings and employment histories, which can directly affect reservation wages and thereby the selection process in the current period. Similarly, persistent earnings shocks or mean reversion may directly influence the current period’s offered wage, again linking recent wages to labor market participation. 

To address these dynamic channels within our identification strategy, we exploit detailed information on each individual’s employment and wage history.
Specifically, we use the earliest observed wage as our main instrument, referring to it as the \textit{initial wage}. This choice mitigates concerns that contemporaneous or short-lagged wages could directly affect current employment decisions.\footnote{In Appendix \ref{appx:robustness}, we show that our results are robust when considering only wages from 2011 or earlier (i.e., at least five years back) as valid instruments.} When constructing the instrument, we prioritize full-time earnings from the earliest available date; if an individual has never worked full-time, we instead use their earliest part-time earnings.\footnote{Ancillary analyses in Appendix \ref{appx:robustness} confirm that our results are robust when restricting the instrument to only past full-time wages.} Finally, we require that the initial wage be observed in 2015 or earlier, i.e., at least two periods before the current observation $t-2$, to ensure that the conditional independence assumption underlying our identification is plausibly satisfied.

\FloatBarrier
\begin{figure}[htbp]
\caption{Years of initial wages}
\label{fig:hist_earl_year}
\centering
\begin{threeparttable}
  \centering
  \vspace{-1em}
  \includegraphics[width=0.65\linewidth]{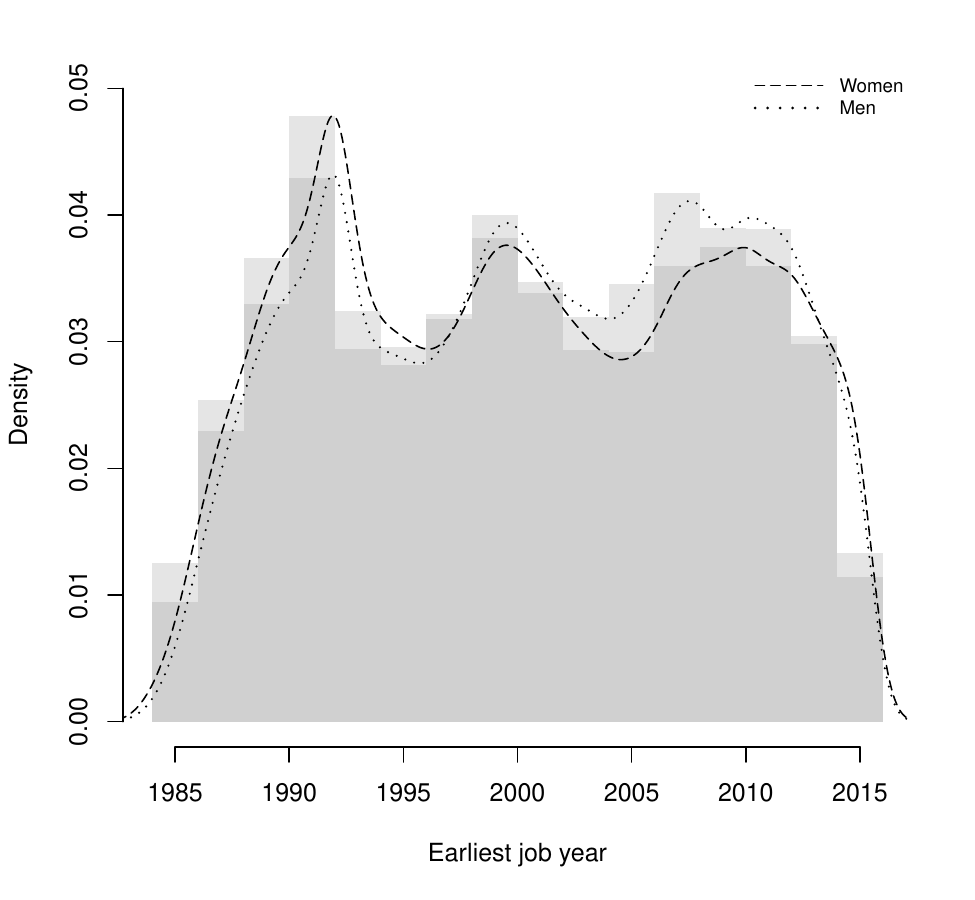}
\begin{tablenotes} \footnotesize
    \item \textit{Note: The figure shows the histogram and density of the earliest job years, with dashed lines representing women and dotted lines representing men.}
\end{tablenotes}
\end{threeparttable}
\end{figure}

\FloatBarrier
\begin{table}[h] \centering \footnotesize
\begin{threeparttable}
  \caption{OLS Regression Results} 
  \label{tab:OLS_results} 
  \footnotesize
\begin{tabular}{@{\extracolsep{5pt}}lcccc} 
\\[-1.8ex]\hline 
\hline \\[-1.8ex] 
 & \multicolumn{2}{c}{\textit{Women:}} & \multicolumn{2}{c}{\textit{Men:}} \\ 
\cmidrule(lr{0.8em}){2-3}
            \cmidrule(lr{0.8em}){4-5} \\[-1.5ex] 

 Initial Wage (log) & 0.216 & 0.183 & 0.272 &  0.183 \\ \vspace{0.5em}
  &  (0.004) &  (0.004) & (0.004) & (0.002)\\ 
    Age & & -0.012 & & -0.011\\ \vspace{0.5em}
  & & (0.000) & & (0.000) \\
 Education Level & & 0.197 & & 0.250\\ \vspace{0.5em}
  & & (0.004) & & (0.002) \\ 
   Workplace & & 0.078 & & 0.205 \\ \vspace{0.5em}
   & & (0.004) & & (0.003)\\ 
 Total Experience & &  0.018 & & 0.019\\ \vspace{0.5em}
 & & (0.000) & & (0.000)\\ 
    Earliest Job Difficulty & & 0.029 & & 0.066 \\ \vspace{0.5em}
  & & (0.003) & & (0.002) \\ 
 Constant & 3.580 & 3.327 & 3.537 & 3.017 \\ \vspace{0.5em}
  & (0.016) & (0.021) & (0.010) & (0.012) \\ 
\hline \\[-1.8ex] 
Observations & \multicolumn{2}{c}{49,876} & \multicolumn{2}{c}{103,703}\\ 
R$^{2}$ & 0.103 & 0.130 & 0.058 & 0.285 \\ 
\hline 
\hline \\[-1.8ex] 
\end{tabular} 
   \begin{tablenotes}[flushleft] \footnotesize
	\item \textit{Note: The dependent variable is the log of full-time wages in 2017. The regression sample includes only individuals who were in full-time employment in 2017.}
    \end{tablenotes}
  \end{threeparttable}
\end{table}

Table \ref{tab:descript_instr} summarizes key characteristics of the instrumental variable, which captures each individual’s earliest observed wage in the dataset. On average, men’s initial wages are higher than women’s, at approximately 60 and 50 euros per day, respectively. The instruments originate from an average age of 22 and reach back around 16 years for both genders. Moreover, 98\% of men’s and 93\% of women’s instruments are based on full-time wages, reflecting the higher incidence of part-time employment among women. Figure \ref{fig:hist_earl_year} displays the distribution of the instrument years, indicating that women’s wage histories extend slightly further back in time than men’s.

In Table \ref{tab:OLS_results}, we assess the relevance of the instrument by regressing log full-time wages observed in 2017 on the initial wage variable. We additionally control for age, educational attainment, workplace region, and total experience, all measured as of 2017, as well as the difficulty level of the earliest job. Total experience is calculated as the sum of full-time experience and half the part-time experience (i.e., one year of part-time work counts as 0.5 years). The coefficient on the initial wage is strongly positive and statistically significant for both men and women, underscoring the predictive power of early labor market conditions for later-life earnings. 

\subsection{Implementation Details} \label{ssec:empirical_model}

We implement quantile regression using three different approaches. First, we apply our proposed IV-based selection correction method. Second, we consider a commonly used selection correction approach that relies on the MAR assumption. Third, we estimate quantile regressions based solely on the observed full-time wages, without correcting for selection. These three methods were compared in detail in the simulation study presented in Section \ref{sec:simulation}.

\paragraph{IV-based selection correction} We apply our three-stage procedure described in Section \ref{ssec:estimation} to estimate selection-corrected wage quantiles separately for men and women. As outlined in Algorithm \ref{alg:est}, the first two stages involve estimating the constrained inverse selection probabilities using the initial wage (log) as an instrument, while controlling for the same covariates $X$ as in Table \ref{tab:OLS_results}. In the third stage, we use the estimated weights to recover the selection-corrected conditional quantile function of the latent outcome, based on the observed full-time wages. This estimation is carried out conditional on both the covariates $X$ and the instrument, in line with Equation \eqref{eq:main_model}. For unconditional quantile estimation, no covariates are included in the third stage. All estimations are conducted separately by gender. Since our analysis focuses on selection into full-time employment, we define $D = 1$ when full-time wage information is observed. Wages of part-time workers and the absence of labor earnings among non-employed individuals are treated as unobserved, corresponding to $D = 0$.

\paragraph{Comparison to MAR correction and uncorrected estimates} A common approach assumes that, conditional on covariates, data are missing at random (MAR), as described in Section~\ref{sec:simulation}. Under this assumption, selection depends only on observable characteristics and not on unobserved outcomes. We implement the MAR-based selection correction using a two-step procedure: first, we estimate inverse probability weights via a probit model; second, we use these weights to estimate the quantile functions. Following prior studies that apply imputation-based methods to address selection, such as \citet{OliPet_08} and \citet{Blau_etal_2021}, we use the same set of variables as in the IV correction but exclude the instrument (initial wage) and its related variable (earliest job difficulty) from both stages of the MAR correction method. For comparison, we also report uncorrected estimates based solely on individuals with observed wages. These are obtained from standard quantile regressions using the same covariates as in the MAR specification. Although straightforward to implement, such estimates may be biased if sample selection is non-random.

\section{Results}\label{sec:results}
We begin by presenting the overall patterns in the unconditional wage quantiles of men and women, comparing our IV-based selection-corrected estimates to both uncorrected estimates and those derived under the MAR assumption. We then turn to the conditional wage quantiles, which account for gender differences in education, experience, and other relevant covariates. These analyses allow us to assess how selection operates across the wage distribution within and between groups. Finally, we quantify the magnitudes of the selection effects and their implications for the gender wage gap across three selected quantiles.

\subsection{Selection Effects in the Unconditional Wage Quantiles}\label{ssec:distrib_uncond}

\begin{figure}[htbp]
    \caption{Quantile Selection Effects (Unconditional)}
    \label{fig:quantiles_uncond2015}
  \begin{threeparttable}
    \centering
    \includegraphics[width=1\linewidth]{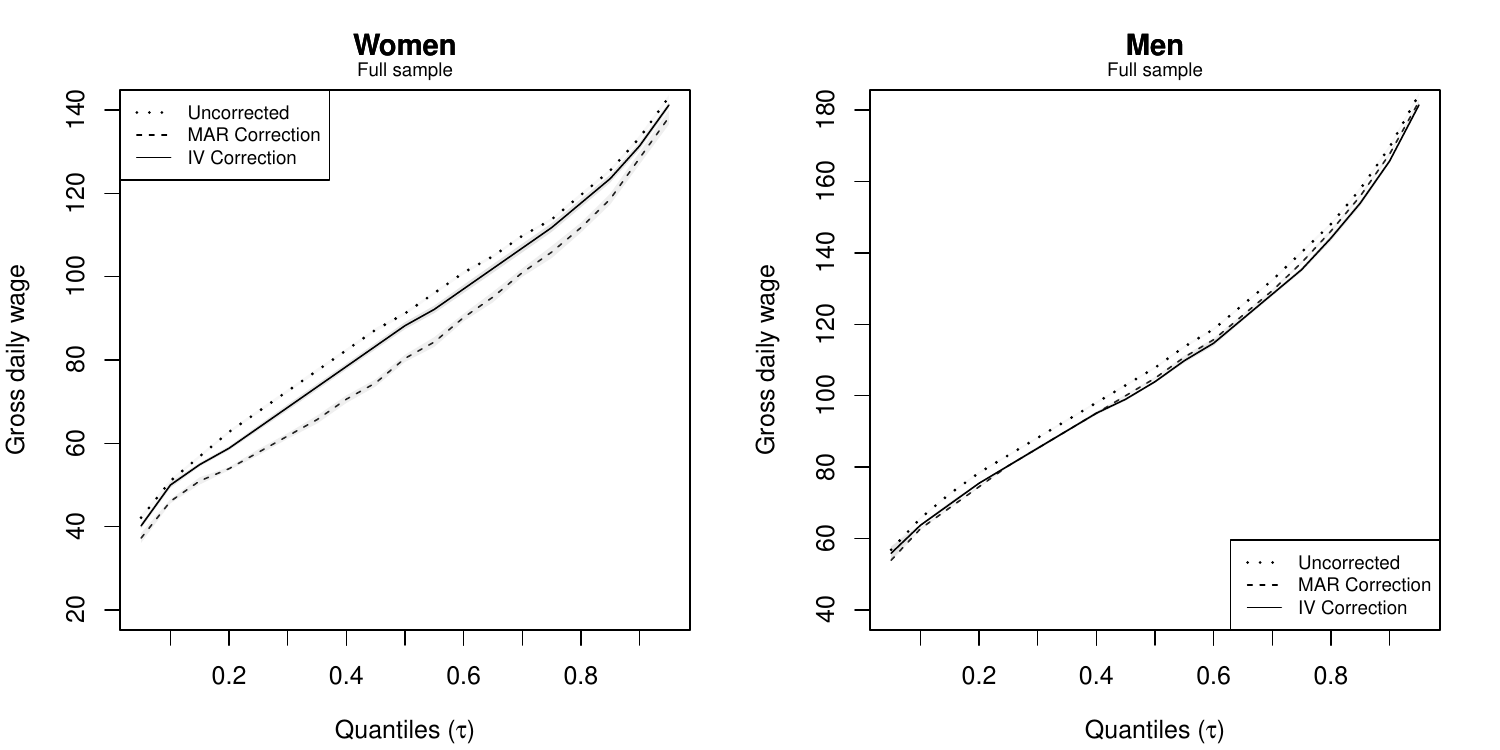}
    \begin{tablenotes}[flushleft] \footnotesize
        \item \textit{Note: SIAB-R data, 2017. Curves display wage point estimates from unconditional quantile regressions across quantiles, connected by lines for visualization: uncorrected (dotted), MAR correction (dashed), and IV correction (solid). The IV correction uses initial log wage as an instrument and controls for education, age, workplace region, total experience, and earliest job difficulty in the first two stages. The MAR correction uses the same controls except earliest job difficulty. The final-stage regressions include no covariates for all methods. Shaded areas represent 95\% pointwise confidence intervals.}
    \end{tablenotes}
    \end{threeparttable}
\end{figure}

The results document sizable and statistically significant positive selection effects for both men and women throughout the wage distribution. Figure \ref{fig:quantiles_uncond2015} displays estimated wage quantiles for men and women, comparing our IV-based selection-corrected estimates with both the uncorrected and MAR-corrected counterparts. The corresponding uncorrected and IV-corrected wage estimates at the 25th, 50th, and 75th percentiles, along with the implied gender wage gaps, are presented in the first panel of Table~\ref{tab:cond_quantiles_edu}.

The selection-corrected unconditional wage quantiles for both women and men lie consistently below the empirical quantiles of wages of observed full-time workers. This indicates positive selection into full-time employment for both genders: individuals who work full-time tend to have higher potential wages than those who do not. As a result, the observed distribution overstates overall wage levels by disproportionately reflecting high-earners. 

Among women, the selection effect is strongest at the lower quantiles and around the median, gradually tapering off toward the upper tail. For example, the empirical median daily wage for women is approximately 91 euros, while the IV-based selection-corrected estimate is about 88 euros, reflecting a 3.2\% reduction. At the 25th percentile, the correction lowers wages by 5.8\%, whereas at the 75th percentile the effect is more modest at 1.7\%. For men, the magnitude of selection is more uniform across the distribution, averaging around 3.5\%. This pattern of positive selection echoes classic insights from \citet{heckman1979sample} and empirical findings such as those by \citet{wang_etal}. It is consistent with the Roy model of labor supply, according to which individuals with higher potential earnings are more likely to select into full-time work, while those with lower potential wages tend to work part-time or remain non-employed. Since both men and women are positively selected, but to different degrees across the distribution, the gender wage gap tends to increase below the median and decrease above it. The findings generally align with and refine the distributional evidence in \citet{arellano2017}, who also find positive selection for both genders but with a larger bias for men in the UK. Our German data reveal that selection among women is particularly influential in the lower half of the distribution, which helps explain why the median gap barely changes while the 25th-percentile gap widens.

Comparing the MAR and IV correction results, we find that both correction methods indicate positive selection for women, with stronger effects in the MAR correction. For men, both approaches imply positive selection effects of similar magnitude. This discrepancy arises because the MAR approach assumes that selection depends only on observed characteristics, ignoring unobserved factors that jointly affect wages and labor supply decisions. The problem is amplified for women, who exhibit substantially higher rates of non-employment and part-time work. As shown in the Monte Carlo simulations in Section \ref{sec:simulation}, the MAR approach performs well only when selection is driven solely by observables, but fails when unobserved heterogeneity plays a role. Consequently, the MAR correction tends to over-adjust in the presence of unobserved selection, producing biased estimates, as illustrated in Figure~\ref{fig:quantiles_uncond2015}.

\subsection{Selection Effects in the Conditional Wage Quantiles}\label{ssec:quantiles}

\begin{figure}[htbp]
\caption{Conditional wage quantiles by education groups}
\label{fig:plot_by_edu}
  \begin{threeparttable}

	\begin{subfigure}{1\textwidth}
 \centering
 \vspace{-0.5em}
 		\caption{Low Education}
            \includegraphics[width=.87\linewidth]{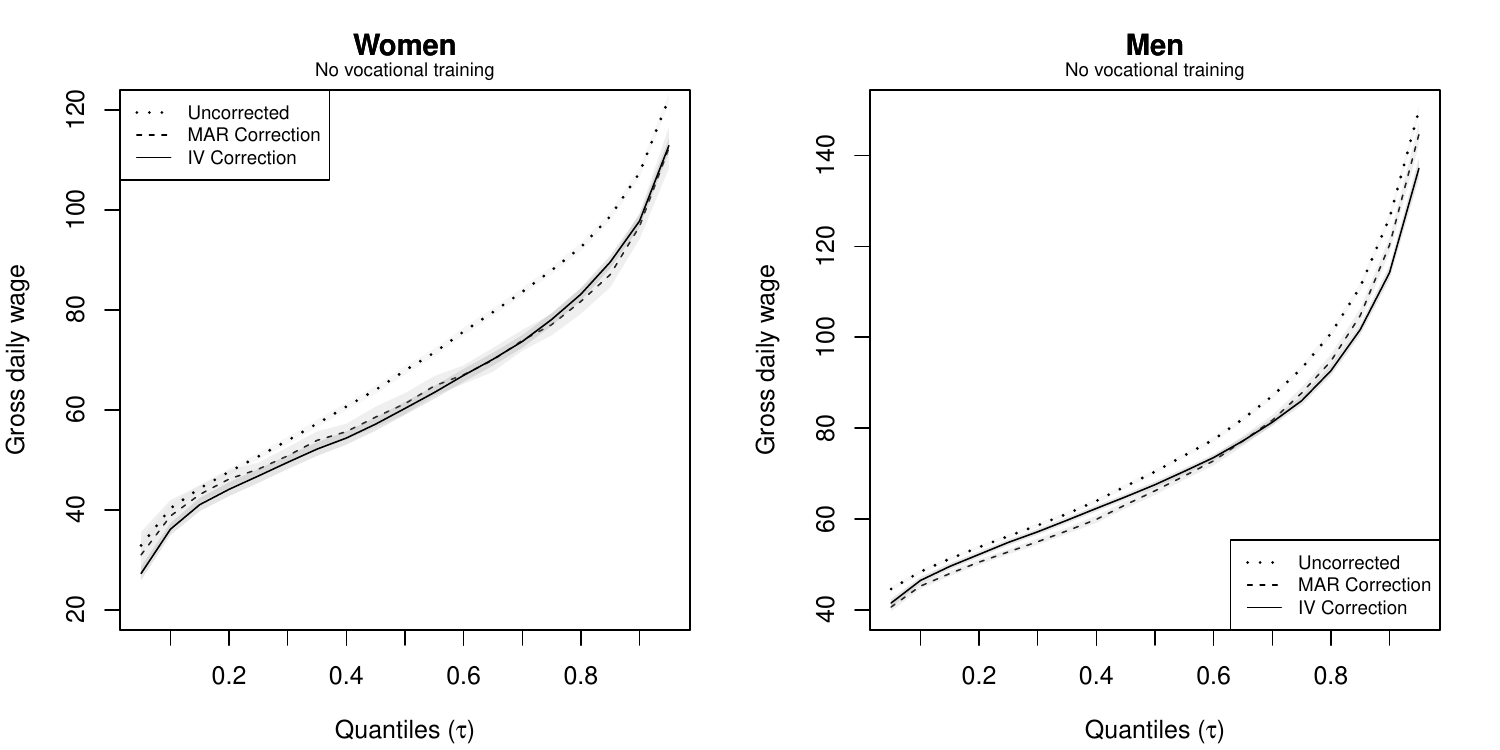}
	\end{subfigure}
 \vspace{-0.3em}
	\begin{subfigure}{1\textwidth}
		\centering
        \vspace{-0.5em}
		\caption{Middle Education}

		\includegraphics[width=.87\linewidth]{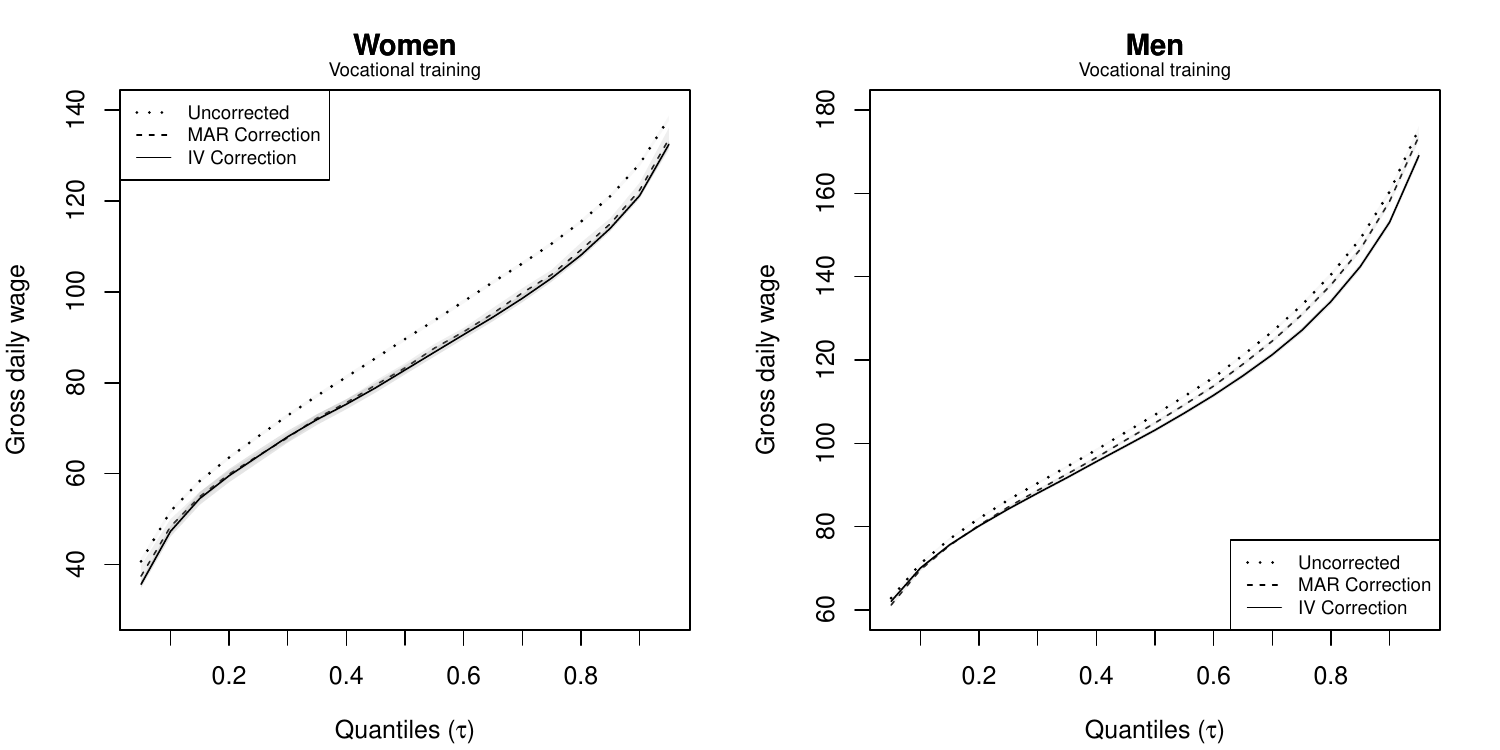}
	\end{subfigure}
\vspace{-0.3em}
 	\begin{subfigure}{1\textwidth}
		\centering
        \vspace{-0.5em}
		\caption{High Education}
		\includegraphics[width=.87\linewidth]{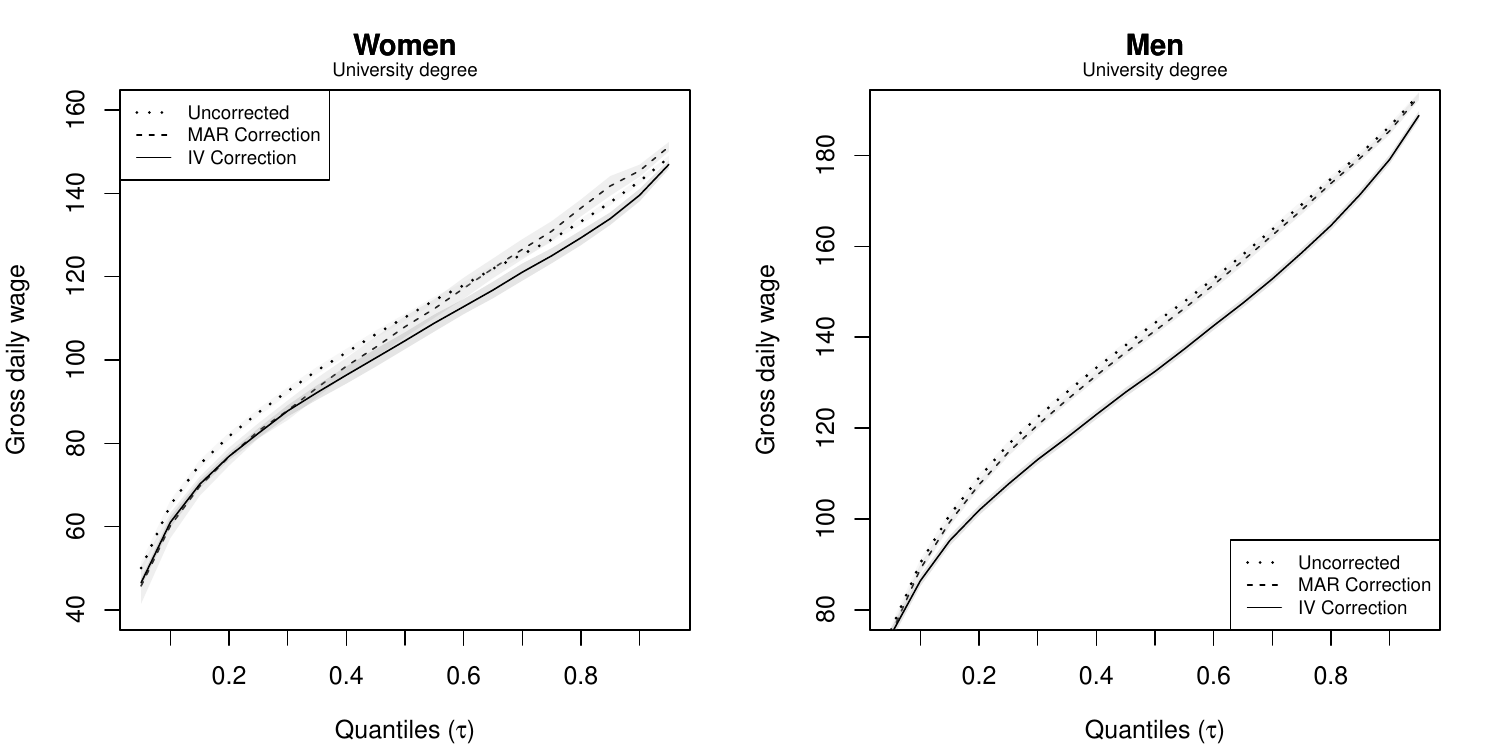}
	\end{subfigure}
    \vspace{-1em}
	\begin{tablenotes}[flushleft] \footnotesize
	    \item \textit{Note: SIAB-R data, 2017. Curves show wage point estimates from unconditional quantile regressions across quantiles, connected by lines for visualization: uncorrected (dotted), MAR correction (dashed), and IV correction (solid). The IV correction uses initial log wage as an instrument and controls for education, age, workplace region, total experience, and earliest job difficulty. The MAR correction uses the same controls, excluding the instrument and earliest job difficulty. The uncorrected estimates include the same controls as in MAR. Shaded areas represent 95\% pointwise confidence intervals.}
	\end{tablenotes}
\end{threeparttable}
\end{figure}

Unconditional wage analyses offer only limited insights, as they do not control for differences in observable characteristics like education, experience, or workplace factors. This can confound true wage disparities with compositional differences across groups. To address this, we turn to conditional quantile regression, which allows us to examine how wages vary across the distribution while accounting for important covariates. Thus, this method captures heterogeneous effects and selection patterns that are often hidden in mean-based or unconditional approaches, providing a more detailed understanding of wage inequality.

Figure \ref{fig:plot_by_edu} displays the conditional wage quantiles by gender and education level, comparing uncorrected estimates based on observed full-time wages with selection-corrected estimates obtained under the MAR assumption and our IV-based method. For women, both correction methods yield broadly similar results across most of the distribution, except at the upper quantiles among the highly educated, where the MAR correction indicates a small negative selection effect beyond the 60th percentile, while the IV-based approach still points to positive selection. For men, selection patterns differ by education level. Among those with low or medium education, differences between the MAR and IV-based methods are relatively minor and occur primarily in the upper parts of the wage distribution. The MAR correction indicates little to no selection effect, while our IV-based method reveals a substantial and statistically significant positive selection.

Overall, wage estimates for women based on the IV method are consistently lower than the uncorrected values across all education groups, with the largest corrections observed among low-educated women and smaller adjustments among the highly educated. This pattern suggests that unobserved characteristics strongly influence selection into full-time employment, especially among lower-educated women. Women in these groups who are observed working full-time appear to have unobserved traits associated with higher wages, leading to a positively selected sample. The IV correction adjusts for this by providing a more representative picture of the full population, including women not observed in full-time work who likely earn lower wages due to their unobserved characteristics.

Men exhibit a distinctly different pattern. For low-educated men, selection effects are modest and become even smaller among those with middle education. By contrast, highly educated men experience substantial positive selection, suggesting that those with higher earning potential are disproportionately represented among full-time workers. This aligns with a setting where full-time participation among highly educated men reflects stronger labor market attachment and career-oriented incentives rather than financial necessity. For this group, employment decisions are likely influenced by unobserved traits such as ambition, productivity, or access to high-return jobs that amplify the benefits of continuous full-time work. In contrast, for less educated men, participation decisions may be driven more by short-term income needs or local labor demand conditions, making selection less sensitive to unobserved earning potential.

While the IV-based and MAR corrections yield broadly similar results for low- and medium-educated women and men, the two approaches diverge most clearly at the top of the wage distribution among the highly educated. This pattern suggests that unobserved factors such as ambition or career orientation play a stronger role in shaping high-end wages, particularly for highly educated men and women. In this group, the MAR correction misestimates selection effects because it fails to account for selection driven by such unobservables, whereas the IV-based approach corrects for them more effectively.

\subsection{Gender wage gaps by education and wage quantile}\label{ssec:gendergap}

Table \ref{tab:cond_quantiles_edu} reports predicted daily wages for men and women and the corresponding wage gaps at selected quantiles (25th, 50th, and 75th), setting education to one of three levels (low, middle, high) while averaging over all other covariates. The first panel of the table additionally presents the unconditional results, providing a baseline comparison before adjusting for observable characteristics. We compare estimates that do not correct for selection and selection corrected estimates using the proposed IV method. A central and straightforward observation is that men consistently earn higher wages than women across all quantiles and education groups, indicating a persistent gender wage gap that holds regardless of educational attainment, quantiles, and estimation method. However, there exist interesting differences between the groups. 

\FloatBarrier
\begin{table}[h!]
    \centering
    \caption{The distribution of the gender wage gap by education}
    \label{tab:cond_quantiles_edu}
    \begin{threeparttable}
    \begin{adjustbox}{width=0.95\linewidth,center}
    \begin{tabular}{l *{9}{c}}
           & \multicolumn{3}{c}{Uncorrected} & \multicolumn{3}{c}{IV selection correction} \\
       
            \cmidrule(lr{0.8em}){2-4}
            \cmidrule(lr{0.8em}){5-7}
            
        & \multicolumn{1}{c}{25th} & \multicolumn{1}{c}{50th} & \multicolumn{1}{c}{75th} & \multicolumn{1}{c}{25th} & \multicolumn{1}{c}{50th} & \multicolumn{1}{c}{75th}\\
    
                     \cmidrule(lr{0.8em}){2-4}
            \cmidrule(lr{0.8em}){5-7}
             \multicolumn{2}{l}{\textbf{Unconditional}} &   &     &  \\
                 \multirow{2}{*}{Men} & 83.33 & 107.84 & 140.20 & 80.39 & 103.92 & 135.29\\
                  & {\scriptsize[83.06,83.60]} & {\scriptsize[107.55,108.13]} & {\scriptsize[139.74,140.65]} & {\scriptsize[80.08,80.71]} & {\scriptsize[103.47,104.37]} & {\scriptsize[134.54,136.05]} \\
              \multirow{2}{*}{Women} & 67.65 & 91.18 & 113.73 & 63.73 & 88.24 & 111.76\\
              & {\scriptsize[67.24,68.06]} & {\scriptsize [90.77,91.58]} & {\scriptsize [113.42,114.03]} & {\scriptsize[62.96,64.50]} & {\scriptsize[87.50,88.97]} & {\scriptsize[110.99,112.54]} \\
                    \cmidrule(lr{0.8em}){2-4}
            \cmidrule(lr{0.8em}){5-7}
            $\%$ Difference  & 18.82 & 15.45 & 18.88 & 20.73 & 15.09 & 17.39\\
                        \cmidrule(lr{0.8em}){2-4}
            \cmidrule(lr{0.8em}){5-7}
         \multicolumn{2}{l}{\textbf{Low Education}} &   &     &  \\
                 \multirow{2}{*}{Men} & 56.26 & 70.44 & 93.15 & 54.88 & 67.57 & 86.02\\
                  & {\scriptsize[55.83,56.68]} & {\scriptsize[69.90,70.99]} & {\scriptsize[92.37,93.93]} & {\scriptsize[54.27,55.49]} & {\scriptsize[66.91,68.23]} & {\scriptsize[85.12,86.94]}\\
              \multirow{2}{*}{Women} & 50.71 & 67.90 & 87.94 & 46.80 & 60.32 & 78.10\\
              & {\scriptsize[50.04,51.40]} & {\scriptsize[67.08,68.74]} & {\scriptsize[87.16,88.73]} & {\scriptsize[45.44,48.21]} & {\scriptsize[59.04,61.62]} & {\scriptsize[76.84,79.37]}\\
                    \cmidrule(lr{0.8em}){2-4}
            \cmidrule(lr{0.8em}){5-7}
            $\%$ Difference & 9.86 & 3.60 & 5.59 & 14.72 & 10.73 & 9.22\\
                        \cmidrule(lr{0.8em}){2-4}
            \cmidrule(lr{0.8em}){5-7}
        \multicolumn{2}{l}{\textbf{Middle Education}}     &  &  &\\
                 \multirow{2}{*}{Men} & 86.44 & 106.85 & 133.13 & 84.25 & 103.17 & 127.07\\
                  & {\scriptsize[86.19,86.69]} & {\scriptsize[106.55,107.16]} & {\scriptsize[132.72,133.55]} & {\scriptsize[83.87,84.63]} & {\scriptsize[102.70,103.65]} & {\scriptsize[126.44,127.70]}\\
                \multirow{2}{*}{Women} & 68.23 & 89.61 & 110.59 & 63.85 & 82.81 & 103.01\\
                & {\scriptsize[67.83,68.63]} & {\scriptsize[89.18,90.06]} & {\scriptsize[110.18,111.00]} & {\scriptsize[62.46,65.27]} & {\scriptsize[81.83,83.79]} & {\scriptsize[102.23,103.80]}\\
                    \cmidrule(lr{0.8em}){2-4}
            \cmidrule(lr{0.8em}){5-7}
            $\%$ Difference & 21.07 & 16.13 & 16.93 & 24.21 & 19.74 & 18.93\\
                            \cmidrule(lr{0.8em}){2-4}
            \cmidrule(lr{0.8em}){5-7}
        \multicolumn{2}{l}{\textbf{High Education}}  &  &  &\\
                 \multirow{2}{*}{Men} & 116.47 & 143.16 & 169.18 & 107.70 & 132.48 & 158.57\\
                  & {\scriptsize[115.73,117.21]} & {\scriptsize[142.43,143.89]} & {\scriptsize[168.38,169.99]} & {\scriptsize[106.70,108.70]} & {\scriptsize[131.50,133.46]} & {\scriptsize[157.53,159.63]}\\
                \multirow{2}{*}{Women}  & 87.38 & 110.24 & 128.90 & 82.42 & 104.61 & 125.00\\
                & {\scriptsize[86.47,88.31]} & {\scriptsize[109.39,111.09]} & {\scriptsize[128.19,129.61]} & {\scriptsize[80.78,84.08]} & {\scriptsize[103.14,106.10]} & {\scriptsize[123.32,126.70]}\\
                    \cmidrule(lr{0.8em}){2-4}
            \cmidrule(lr{0.8em}){5-7}
            \cmidrule(lr{0.8em}){8-10}
            $\%$ Difference & 24.97 & 23.00 & 23.81 & 23.47 & 21.04 & 21.17\\
        \cmidrule(lr{0.8em}){2-7} 
    \end{tabular}
    \end{adjustbox}
    \end{threeparttable}
    \begin{threeparttable} 
    \vspace{-1em}
        \begin{tablenotes}[flushleft]
        \footnotesize
        \item \textit{Note: SIAB-R data, 2017. The table presents quantile regression estimates for the 0.25, 0.50, and 0.75 quantiles. The IV correction uses initial log wage as an instrument and controls for education, age, workplace region, total experience, and earliest job difficulty in the first two stages. In the first panel (unconditional) quantile regressions include no covariates. In the remaining panels (conditional), the IV correction carries the same variables to the third stage. The uncorrected estimates use the same controls, excluding the instrument and earliest job difficulty. Conditional estimates fix controls at their median within each education group. 95\% confidence intervals are shown in square brackets.}
        \end{tablenotes}
    \end{threeparttable}
   
\end{table}

The unconditional estimates in the first panel, corresponding to Figure \ref{fig:quantiles_uncond2015}, capture the differences in wage distributions without adjusting for covariates. At the median, men earn about 108 euros per day, compared to 91 euros for women, implying a raw wage gap of 15.5\%. After IV-based selection correction, median wages fall to 104 euros for men and 88 euros for women, leaving the gap virtually unchanged. At the lower quantile, however, the correction increases the wage gap by roughly 2 percentage points to 20.7\%, while at the upper quantile it reduces the gap by about 1.5 percentage points to 17.4\%.

In the second panel, where education is set to no vocational training (low education), we find strong positive selection for women and significant but less pronounced positive selection for men. The uncorrected daily wages for men are 56 euros, 70 euros, and 93 euros at the 25th, 50th, and 75th percentiles, respectively. For women, the corresponding figures are 51, 68, and 88 euros. After applying the IV-based selection correction, median wages fall to 68 euros for men and 60 euros for women, increasing the gender wage gap from 3.6\% to 10.7\%. At the 25th percentile, strong positive selection for both genders widens the gap by 4.8 percentage points, whereas at the 75th percentile the effect is smaller, raising the gap by only 3.6 percentage points.

Among workers with vocational training (middle education), the gender wage gap widens across all quantiles, although the changes are less dramatic than for the low-education group. At the median, men’s wages fall from 107 to 103 euros and women’s from 90 to 83 euros, widening the gender wage gap by 3.6 percentage points to 19.7\%. At the lower end of the distribution, the gap increases by 3.1 percentage points to 24.2\%, while at the 75th percentile, the correction adds only 2 percentage points.

For university graduates, selection patterns differ sharply between men and women. Women with higher education are only marginally affected by selection, whereas highly educated men experience the strongest positive selection in the sample. At the upper quantiles, men’s daily wages are about 11 euros lower after correction, compared to a smaller decline of 4–5 euros for women. These opposing effects even reduce the gender wage gap at the top of the distribution, by roughly 1–2 percentage points.

Relative to the results in \citep[e.g.,][]{Blau_etal_2021}, our IV-based estimates attribute more of the observed female advantage among full-time workers to selection on unobservables, especially at the bottom of the distribution. In this sense, our results caution against interpreting MAR-corrected improvements as purely structural convergence: once non-ignorable selection is accommodated, distributional gaps at the lower tail are larger, while top-end gaps can be smaller due to stronger male selection.

Taken together, our estimates both reconcile and extend existing literature. In addition to documenting positive selection throughout the wage distribution, we uncover two distinctive patterns: (i) pronounced selection at the lower end for women, especially those with less education, which widens the gender gap in this segment; and (ii) strong positive selection among highly educated men at the top of the distribution, which narrows the gender wage gap at upper quantiles. These findings rationalize why selection-corrected gender wage gaps can widen, as shown by \citet{MaasWang_19}, while in other cases they may narrow at the upper end of the distribution, consistent with \citet{arellano2017}.

\section{Conclusion}\label{Sec:Conclusion}

In this paper, we propose a novel estimation strategy to recover the selection-corrected wage distribution and to quantify the full distribution of the gender wage gap. Identification is obtained from exogenous variation that shifts latent wages but does not, conditional on these latent and other observed variables, provide additional information on the selection mechanism. The approach does not impose parametric restrictions on the selection probability and, within the Roy model, can be interpreted as a rank invariance condition on conditional reservation wages.

We formally derive an estimation strategy for selection-corrected quantile regression when the outcome is selectively observed and selection depends on unobservables. We use inverse probability weighting in combination with an instrumental variable approach, thereby relaxing the MAR assumption that selection depends only on observables. We use shape constraints via cone projection and B-spline approximations in the estimation of inverse probabilities, which stabilize the weights and improve inference. 

Using administrative social security data from Germany, we apply the proposed estimation strategy to derive selection-corrected wage distributions for men and women in full-time employment. The results of our analysis highlight the critical role of selection in shaping the gender wage gap. We document substantial selection at the lower tail for women, amplifying gaps specifically among the less educated, and strong positive selection among highly educated men, compressing gaps at the top. Despite these differences in the selection pattern, men consistently earn higher full-time wages than women across all quantiles and education groups. Thus, the gender wage gap persists regardless of educational attainment, quantiles, and selection patterns. 

\bibliography{bibliography}

\newpage

\appendix

\newgeometry{top=2in}

\begin{center}

{\LARGE Online Appendix to `Quantile Selection in the\\[1em] Gender Pay Gap'}\\[6em]
\end{center}

{\Large \textbf{Table of Contents}}\\[0.1em]

\setcounter{section}{0}
\setlength{\cftsecnumwidth}{2em}

\startcontents[sections]
\printcontents[sections]{l}{1}{\setcounter{tocdepth}{2}}

\thispagestyle{empty}

\restoregeometry
\clearpage
\setcounter{page}{1}
\pagenumbering{arabic}
\renewcommand{\thesection}{A\arabic{section}}
\section{Proofs of the Identification Results in Section~\ref{sec:model_identification}}\label{appx:proofs}
\normalsize
\begin{proof}[\textsc{Proof of Proposition \ref{lem:identification}}]
For any $\tau \in (0,1)$, it is well known (see \cite{KoenkerBassett}) that the quantile regression coefficient $\theta_\tau$ minimizes the check function as given in Equation \eqref{eq:QR_equation}, i.e.,
\[
\theta_\tau = \argmin_{\theta \in \mathbb{R}^{d_z}} \, 
\Ex\!\left[\left(Y^*-Z^\top \theta\right)\left(\tau - \1\{Y^* < Z^\top \theta\}\right)\right].
\]
Multiplying and dividing by $\PP\left(D=1 \mid Y^*,X\right)$ gives
\begin{align*}
\theta_\tau 
&= \argmin_{\theta \in \mathbb{R}^{d_z}} 
\Ex\!\left[\left(Y^*-Z^\top\theta\right)\left(\tau - \1\{Y^* < Z^\top \theta\}\right)
\frac{\PP\left(D=1 \mid Y^*,X\right)}{\PP\left(D=1 \mid Y^*,X\right)}\right].
\end{align*}
By the exclusion restriction on $W$, we have $\PP\left(D=1 \mid Y^*,X,W\right) = p\left(Y^*,X\right)$, so that
\begin{align*}
\theta_\tau 
&= \argmin_{\theta \in \mathbb{R}^{d_z}} 
\Ex\!\left[\left(Y^*-Z^\top\theta\right)\left(\tau - \1\{Y^* < Z^\top \theta\}\right)
\frac{\PP\left(D=1 \mid Y^*,X,W\right)}{\PP\left(D=1 \mid Y^*,X\right)}\right].
\end{align*}
Applying the law of iterated expectations and noting that 
$Y=Y^*$ whenever $\Delta=1$, with $p\left(Y^*,X\right)=p\left(Y,X\right)$ on this support, we obtain
\begin{align*}
\theta_\tau 
&= \argmin_{\theta \in \mathbb{R}^{d_z}} 
\Ex\!\left[\left(Y-Z^\top\theta\right)\left(\tau - \1\{Y < Z^\top \theta\}\right) 
\frac{\Delta}{p\left(Y,X\right)}\right] \\
&= \argmin_{\theta \in \mathbb{R}^{d_z}} 
\Ex\!\left[\frac{\Delta}{p\left(Y,X\right)} \, \rho_\tau\left(Y - Z^\top \theta\right)\right],
\end{align*}
which proves the claim.
\end{proof}

\begin{proof}[\textsc{Proof of Theorem \ref{prop:1}}]

    We first introduce the linear operator $T$ by $T:f \mapsto \Ex[D f(Y^*,X) \mid W,X]$ for any square integrable function $f$. Then, the conditional mean equation in \eqref{cond:eq}, that is, $\Ex[D g(Y^*,X) \mid W,X] = 1$, can be written as the operator equation $Tg = 1$.
Furthermore, the quantile regression coefficient $\theta_\tau$ minimizes a population criterion as in Equation~\eqref{eq:theta_tau} which can be rewritten as
    \[
    \Ex\left[g(Y^*,X) m_\tau(Y^*,X)\right], \quad \text{with}\quad m_\tau(Y^*,X) := \Ex\left[D \rho_\tau\left(Y^* - Z^{\top} \theta_\tau\right) \bigm| Y^*,X\right].
    \]
This population criterion thus immediately implies uniqueness of the  quantile regression coefficient $\theta_\tau$.   
    
    We can now apply Lemma 3.1 in \cite{severini_tripathi2012}: Identification of the population criterion $\Ex\left[g(Y^*,X) m_\tau(Y^*,X)\right]$ requires only that $m_\tau$ belongs to the orthogonal complement (in the $L^2$ sense) of the operator $T$. Equivalently, there exists a function $\mu_\tau$ such that $T^*\mu_\tau = m_\tau$, where the adjoint operator is defined as $T^*: \phi \mapsto \mathbb{E}[D\phi(W,X)\mid Y^*, X]$. This condition can be written as
    \begin{align*}
        \Ex\left[D \mu_\tau (W,X)\mid Y^*,X\right] = \Ex\left[D \rho_\tau\left(Y^* - Z^{\top} \theta_\tau\right) \mid Y^*,X\right].
    \end{align*}
    Using the selection equation $D = \1\{V \leq p(Y^*,X)\}$ and the conditional independence $V\independent W\mid (Y^*, X)$, we can factor out the probability of selection $p(Y^*,X) = \Ex[D\mid Y^*,X]$, yielding
    \begin{equation*}
        p(Y^*,X)\Ex\left[\mu_\tau(W,X)\mid Y^*,X\right]=p(Y^*,X)\Ex\left[\rho_\tau\left(Y^* - Z^{\top} \theta_\tau\right) \mid Y^*,X\right].
    \end{equation*}
    By the overlap condition $p(Y^*,X)>0$, we divide both sides by $p(Y^*,X)$ to obtain the equivalent expression
    \begin{equation*}
        \Ex\left[\mu_\tau(W,X)\mid Y^*,X\right]=\Ex\left[\rho_\tau\left(Y^* - Z^{\top} \theta_\tau\right) \mid Y^*,X\right],
    \end{equation*}
which completes the proof.
\end{proof}

\begin{proof}[\textsc{Proof of Corollary \ref{coro:roy_model}}]
Since $R$ conditional on $(Y^*,X)$ is continuously distributed, the conditional cdf. $F_{R|Y^*,X}(\cdot)$ is strictly increasing. Consequently, we may write the Roy selection rule as
\begin{align*}
    D&=\mathbbm1\left\{Y^* \geq R\right\}\\
    &=\mathbbm1\left\{F_{R|Y^*,X}(Y^*) \geq F_{R|Y^*,X}(R)\right\}\\
    &=\mathbbm1\left\{p(Y^*,X) \geq F_{R|Y^*,X}(R)\right\}.
\end{align*}
This corresponds exactly to the general selection model \eqref{selection:mod} with $V=F_{R|Y^*,X}(R)$. Hence, we can apply the proof of Theorem \ref{prop:1}, replacing $V$ by $F_{R|Y^*,X}(R)$ to obtain identification of the quantile regression coefficients $\theta_\tau$. 
\end{proof}

\section{Technical Assumptions and Proof of Theorem~\ref{thm:asymptotic_normality}} \label{appx:thm3proof}
The following conditions are required to establish Theorem~\ref{thm:asymptotic_normality}. We observe a random sample $\{S_i\}_{i=1}^n$, where $S_i = (D_i, Y_i, W_i^\top, X_i^\top)$.
Define the identified set of inverse selection probability functions as $\mathcal{I} := \{g \in L^2(Y,X) : \Ex[Dg(Y,X) \mid W,X] = 1\}$.
Let $H = \Ex[D\phi^J(Y,X) b^K(W,X)^\top]$ and define the quantile score function $\psi_\tau(u) = \tau - \1\{u < 0\}$. Recall that $M_{1\tau g} = \Ex \big[ \Omega_g f_{Y\mid\Omega_g, Z}\left(Z^\top \theta_\tau\right)Z Z^{\top} \big].$
\renewcommand\theassumption{A\arabic{assumption}}

\begin{assumption}
    \label{A_asymptotics2}
    (i) \label{A_full_rank}
    $H$ is of full rank for all $J, K \in \mathbb{N};$
    (ii) \label{A_continuous_F} The conditional distribution function $F_{Y|\Omega_g, Z}$ is absolutely continuous, with continuous densities $f_{Y|\Omega_g, Z}(\xi)$ uniformly bounded away from 0 and $\infty$ at the points $Z_i^\top \theta_\tau, i=1, \ldots, n$, uniformly for $g \in \mathcal{I}$; (iii) \label{A_ZZ_invert}  $\Ex[ZZ^\top]$ is positive definite.
\end{assumption}

Assumption~\hyperref[A_asymptotics2]{\ref{A_asymptotics2}(i)} ensures that the matrix $H$ is invertible and that the projection of the inverse selection functions onto the sets of basis functions considered is unique. Assumption~\hyperref[A_continuous_F]{\ref{A_asymptotics2}(ii)} requires the conditional distribution of the outcome to be smooth and well-behaved at the linear quantile model points. Together, Assumptions \hyperref[A_asymptotics2]{\ref{A_asymptotics2}(ii)} and \hyperref[A_asymptotics2]{\ref{A_asymptotics2}(iii)} imply that the matrix $M_{1\tau g}$ is invertible.

For each $g\in\mathcal I$ recall the definition $\Omega_g = D g(Y, X)$ of the selection weights. Note that by Assumption~\hyperref[A_asymptotics2]{\ref{A_asymptotics2}(i)} is uniquely determined when the function $g$ belongs to the sieve space under consideration. The corresponding estimator is denoted by $\widehat{\Omega} := D \widehat{g}(Y, X)$. From Equation~\eqref{eq:main_model}, $U_\tau = Y^* - Z^\top \theta_\tau$, which reduces to $U_\tau = Y - Z^\top \theta_\tau$ whenever multiplied by $D$, since $Y = Y^*$ holds for $D=1$. We write $\|\cdot\|_\infty$ for the supremum norm and, for any $\delta \in \mathbb{R}^{d_z}$, define
\[
R_g(\delta) := \big(\beta^\top \phi^J(Y,X) - g(Y,X)\big) \, D \, Z^\top \delta \, \psi_\tau(U_\tau),
\]
which captures the weighted approximation remainder from replacing $g(Y,X)$ by its series approximation $\beta^\top \phi^J(Y,X)$.

\begin{assumption}
\label{A_asymptotics1}
    (i) \label{A_rate_J}
      $\sup_{(y,x) \in \mathcal{Y} \times \mathcal{X}}\|\phi^J(y,x)\|=O(\sqrt J)$ such that $J \log(J)/n=o(1)$, and $K=cJ$ for some constant $c \geq 1$;
     (ii) \label{A_approx_error}
    $\sqrt{n}\, \Ex[R_g(\delta)]=o\big(\sqrt{\Var(R_g(\delta))}\big)$ uniformly for $g\in\mathcal I$;
    (iii) \label{A_moments} 
    $\Ex\left[\left(1 - D\phi^J(Y,X)^\top\beta)^2 \mid W,X\right)\right]$ is bounded away from zero and infinity, and $\Ex\left[\left(1-D\phi^J(Y,X)^\top\beta\right)^4\right]< \infty$;
    (iv) \label{A_variance_bound} $\max _{1\leq i\leq n}\left\|\Omega_{gi} Z_i\right\| =o( \sqrt{n})$ almost surely, uniformly for $g \in \mathcal{I}$;
    (v) \label{A_sup_Omega}
    $\sup_{g\in\mathcal I}\|\widehat{g}-g\|_\infty=o_p(1)$.
\end{assumption}

Assumption \ref{A_asymptotics1} captures standard regularity conditions for establishing inference for semi-/nonparametric series estimators (see, e.g., \cite{Chen07}, \cite{Koenker2005}, \cite{knight1998}, \cite{hjortpollard}). Assumption~\hyperref[A_rate_J]{\ref{A_asymptotics1}(i)} imposes an upper bound on the growth of the series dimension $J$. This ensures that the series approximation remains well-behaved as the sample size grows. Assumption~\hyperref[A_approx_error]{\ref{A_asymptotics1}(ii)} controls the relative size of approximation errors when $g(Y_i, X_i)$ is approximated by $\beta^\top \phi^J(Y_i,X_i)$, ensuring they vanish as the sample grows. Assumption~\hyperref[A_moments]{\ref{A_asymptotics1}(iii)} guarantees the existence of key moments and that certain conditional variances are positive, which underpins the asymptotic normality of $\widehat{\beta}$. Assumption~\hyperref[A_variance_bound]{\ref{A_asymptotics1}(iv)} ensures that no single observation dominates the weighted sums, effectively satisfying a Lindeberg condition. Finally, Assumption~\hyperref[A_sup_Omega]{\ref{A_asymptotics1}(v)} controls the estimation error of the inverse probabilities and the resulting weights, ensuring that the estimated weights converge uniformly to the true weights (see \cite{chen2018optimal}).

\begin{proof}[\textsc{Proof of Theorem \ref{thm:asymptotic_normality}}]
Consider an objective function
$$
S_n\left(\delta\right)=\sum_{i=1}^n \widehat{\Omega}_i \Big[\rho_\tau\left(U_{\tau i}-Z_i^{\top} \delta / \sqrt{n}\right)-\rho_\tau\left(U_{\tau i}\right)\Big]
$$
where $U_{\tau i}=Y_i-Z_i^{\top} \theta_{\tau}$. The function $S_n\left(\delta\right)$ is convex and is minimized at
$$
\widehat{\delta}=\sqrt{n}\left(\widehat{\theta}_{\tau}-\theta_{\tau}\right)
$$
as we know from Equation~\eqref{eq:theta_hat} that 
\[ \min_{\theta \in \mathbb{R}^{d_z}} \sum_{i=1}^n \widehat{\Omega}_i \, \rho_\tau\!\bigl(Y_i - Z_i^\top \theta\bigr) = \sum_{i=1}^n \widehat{\Omega}_i \, \rho_\tau\!\bigl(Y_i - Z_i^\top \widehat{\theta}_\tau\bigr), \]
which is equivalent to $\sum_{i=1}^n \widehat{\Omega}_i \, \rho_\tau\!\bigl(U_{\tau i} - Z_i^\top \widehat{\delta} / \sqrt{n}\bigr)$.

Following \citet{knight1998}, it can be shown that the asymptotic distribution of $\widehat{\delta}$ is determined by the limiting behavior of the function $S_n\left(\delta\right)$. Using Knight's identity,
$$
\rho_\tau\left(u-v\right)-\rho_\tau\left(u\right)=-v \left(\tau-\1\{u<0\}\right)+\int_0^v\left(\1\{u \leq s\}-\1\{u \leq 0\}\right) d s
$$
we may write
$$
S_n\left(\delta\right)=S_{1 n}\left(\delta\right)+S_{2 n}\left(\delta\right),
$$
where
$$
\begin{aligned}
& S_{1 n}\left(\delta\right)=-\frac{1}{\sqrt{n}} \sum_{i=1}^n \widehat{\Omega}_i Z_i^{\top} \delta \left(\tau-\1\{U_{\tau i}<0\}\right) \\
& S_{2 n}\left(\delta\right)=\sum_{i=1}^n \widehat{\Omega}_i \int_0^{Z_i^{\top} \delta / \sqrt{n}}\left(\1\{U_{\tau i} \leq s\}-\1\{U_{\tau i} \leq 0\}\right) d s =: \sum_{i=1}^n \widehat{\Omega}_i S_{2 n i}\left(\delta\right)
\end{aligned}
$$
and we control each term separately in the following.

\paragraph{The first term}
We can write $S_{1n}\left(\delta\right)$ as 
\begin{align*}
\hspace{-2em}
    S_{1 n}\left(\delta\right)&=-\frac{1}{\sqrt{n}} \sum_{i=1}^n \Omega_{gi} Z_i^{\top} \delta \left(\tau-\1\{U_{\tau i}<0\}\right) -\frac{1}{\sqrt{n}} \sum_{i=1}^n \left(\widehat{\Omega}_i - \Omega_{gi}\right) Z_i^{\top} \delta \left(\tau-\1\{U_{\tau i}<0\}\right) \\
    &=-\left(\frac{1}{\sqrt{n}} \sum_{i=1}^n \Omega_{gi} Z_i^{\top} \delta \psi_\tau(U_{\tau i}) + \frac{1}{\sqrt{n}}\sum_{i=1}^n \left(\widehat{\Omega}_i - \Omega_{gi}\right) Z_i^{\top} \delta \psi_{\tau}(U_{\tau i}) \right)\\
    &=: - (A_1 + A_2),
\end{align*}
where we used the definition $\psi_\tau\left(u\right) = \tau - \1\{u < 0\}$ in the second line.

Consider $A_{1}$. From the definition of $\theta_{\tau}$ in Equation~\eqref{eq:QR_equation}, Proposition~\ref{lem:identification} ensures that $\theta_\tau$ minimizes the expected weighted check-loss function $\Ex\big[\Omega_g \rho_\tau\left(Y - Z^\top \theta\right)\big]$ for all $g\in\mathcal I$. This implies $\Ex\big[\Omega_g Z \psi_\tau \left(U_\tau\right)\big] = 0,$ and, in particular, for some vector $\delta$ and any $g\in\mathcal I$, that 
\[\Ex\left[\Omega_g Z^\top \delta \psi_\tau\left(U_\tau\right)\right] = 0 .\] 

Now consider $A_{2}$. We write $A_{2}$ as
\begin{align*}
   \hspace{-2em} A_{2} = \frac{1}{\sqrt{n}} \sum_{i=1}^n \left(\widehat{\Omega}_i - \Omega_{gi} \right)Z_i^{\top} \delta \psi_\tau\left(U_{\tau i}\right) &= \frac{1}{\sqrt{n}} \sum_{i=1}^n \left(\widehat{\Omega}_i - \widetilde{\Omega}_i + \widetilde{\Omega}_i - \Omega_{gi}\right) Z_i^{\top} \delta \psi_\tau\left(U_{\tau i}\right),
\end{align*}
where $\widetilde{\Omega}_i := \beta^\top \phi^J(Y_i,X_i)D_i$. By writing $\widehat{\Omega}_i = \widehat{\beta}^\top \phi^J(Y_i,X_i)D_i$, we can further write
\begin{align*} 
  \hspace{-2em} A_{2} &= \frac{1}{\sqrt{n}} \sum_{i=1}^n \Big[(\widehat{\beta}-\beta)^\top \phi^J(Y_i,X_i) 
       + \big(\beta^\top \phi^J(Y_i,X_i) -  g(Y_i, X_i)\big)
       D_i Z_i^{\top} \delta \psi_\tau(U_{\tau i}) \Big]\\[6pt]
    &= \sqrt{n}(\widehat{\beta}-\beta)^\top 
       \Bigg( \frac{1}{n}\sum_{i=1}^n D_i\phi^J(Y_i, X_i) Z_i^{\top} \delta \psi_\tau(U_{\tau i}) \Bigg) \\
    &\quad+ \frac{1}{\sqrt{n}}\sum_{i=1}^n 
       \big(\beta^\top \phi^J(Y_i,X_i) -  g(Y_i, X_i)\big) D_i Z_i^{\top} \delta \psi_\tau(U_{\tau i}).
\end{align*}
By Assumption~\hyperref[A_approx_error]{\ref{A_asymptotics1}(ii)}, the last summand vanishes asymptotically. For the first summand we have 
\[
\frac{1}{n}\sum_{i=1}^n \phi^J(Y_i, X_i) D_i Z_i^{\top} \psi_\tau(U_{\tau i}) 
=\Ex\!\left[\phi^J(Y, X) D Z^{\top} \psi_\tau(U_\tau)\right] +o_p(1)=T+o_p(1),
\]
and thus $A_2$ becomes
\begin{align*}
    A_2 = \delta^\top T^\top(H G ^{-1} H^\top)^{-1} H G ^{-1}b^K(W_i,X_i) \frac{1}{\sqrt{n}}\sum_{i=1}^n U_{Ji} + o_p(1).
\end{align*}

Hence, $S_{1n}(\delta)$ becomes
\begin{align*}
    S_{1 n}(\delta)&=-\frac{1}{\sqrt{n}} \delta^\top \sum_{i=1}^n M_{0\tau g}(S_i) +o_p(1),
\end{align*}
where we define $M_{0\tau g}(S_i) :=Z_i\Omega_{gi} \psi_\tau\left(U_{\tau i}\right)+T^\top(H G ^{-1} H^\top)^{-1} H G ^{-1}b^K(W_i,X_i)U_{Ji} $.

In particular, for all $g \in \mathcal{I}$ we obtain that
\begin{align*}
    S_{1n}\left(\delta\right)\xrightarrow{d}  -\delta^\top \mathcal Z_{\tau g}, 
\end{align*}
where $\mathcal Z_{\tau g} \sim \mathcal{N}\left(0,\Ex\left[M_{0\tau g}(S) M_{0\tau g}^\top(S)\right]\right)$.

\paragraph{The second term} We write the second term for all $g \in \mathcal{I}$ as
\begin{align*}
    S_{2 n}\left(\delta\right)
    &=\sum_{i=1}^n \Omega_{gi} S_{2ni}\left(\delta\right) + \sum_{i=1}^n \left(\widehat{\Omega}_i - \Omega_{gi}\right) S_{2ni}\left(\delta\right) \\
    &=\sum_{i=1}^n \Ex [\Omega_{gi} S_{2 n i}\left(\delta\right)\mid\Omega_{gi}, Z_i]+\sum_{i=1}^n \left(\Omega_{gi} S_{2 n i}\left(\delta\right)-\Ex [\Omega_{gi} S_{2 n i}\left(\delta\right)\mid\Omega_{gi}, Z_i]\right) \\  &+ \sum_{i=1}^n \left(\widehat{\Omega}_i - \Omega_{gi}\right) S_{2ni}\left(\delta\right) \\
    &=: T_1 + T_2 + T_3.
\end{align*}
Consider $T_1$. Under Assumption~\hyperref[A_continuous_F]{\ref{A_asymptotics2}(ii)}, we have for $g \in \mathcal{I}$ \\
$$\begin{aligned}
\sum_{i=1}^n &\Ex[\Omega_{gi} S_{2 n i}\left(\delta\right)\mid\Omega_{gi}, Z_i] \\
& =\sum_{i=1}^n \Omega_{gi} \Ex\left[ \int_0^{Z_i^{\top} \delta / \sqrt{n}}\left(\1\{Y_i \leq Z_i^\top \theta_\tau+s\}-\1\{Y_i \leq Z_i^\top \theta_\tau\}\right) d s \biggm|\Omega_{gi}, Z_i\right]\\
& =\sum_{i=1}^n \Omega_{gi} \int_0^{Z_i^{\top} \delta / \sqrt{n}}\left(F_{Y\mid\Omega_g, Z}\left(Z_i^\top \theta_\tau+s\right)-F_{Y\mid\Omega_g,Z}\left(Z_i^\top \theta_\tau\right)\right) d s \\
& =\frac{1}{\sqrt{n}} \sum_{i=1}^n \Omega_{gi} \int_0^{Z_i^{\top} \delta}\left(F_{Y\mid\Omega_g, Z}\left(Z_i^\top \theta_\tau+t / \sqrt{n}\right)-F_{Y\mid\Omega_g, Z}\left(Z_i^\top \theta_\tau\right)\right) d t \\
& =n^{-1} \sum_{i=1}^n \Omega_{gi} \int_0^{Z_i^{\top} \delta} \sqrt{n}\left(F_{Y\mid\Omega_g, Z}\left(Z_i^\top \theta_\tau+t / \sqrt{n}\right)-F_{Y\mid\Omega_g, Z}\left(Z_i^\top \theta_\tau\right)\right) d t,
\end{aligned}$$
where in the second line we plugged in $U_{\tau i} = Y_i - Z_i^\top \theta_\tau$, and in the third line we used the fact that $\Ex[\1\{Y_i \leq Z_i^\top \theta_{\tau} + s\}\mid\Omega_{gi}, Z_i] =F_{Y\mid\Omega_g,Z}\left(Z_i^\top \theta_{\tau} +s\right).$ The fourth line follows from substitution. Next, by applying a Taylor expansion to the term \(F_{Y\mid\Omega_g, Z}\left(Z_i^\top \theta_\tau + t/\sqrt{n}\right)\), we get
$$\begin{aligned}
\sum_{i=1}^n \Ex[\Omega_{gi} S_{2 n i}\left(\delta\right)\mid\Omega_{gi}, Z_i] & =n^{-1} \sum_{i=1}^n \Omega_{gi} \int_0^{Z_i^{\top} \delta} f_{Y\mid\Omega_g, Z}\left(Z_i^\top \theta_\tau\right) t d t+o_p\left(1\right) \\
& =\left(2 n\right)^{-1} \sum_{i=1}^n \Omega_{gi} f_{Y \mid\Omega_g, Z}\left(Z_i^\top \theta_\tau\right) \delta^{\top} Z_i Z_i^{\top} \delta+o_p\left(1\right) \\
& = o_p\left(\frac{1}{2} \delta^{\top} M_{1\tau g} \delta\right)
\end{aligned}$$
for all $g \in \mathcal{I}$, where we define $M_{1\tau g} := \Ex \big[ \Omega_g f_{Y\mid\Omega_g, Z}\left(Z^\top \theta_\tau\right)Z Z^{\top} \big].$

Now consider $T_2$. We obtain
\begin{equation*}
\Ex|T_2|^2 = \sum_{i = 1}^n \Ex[\left(\Omega_{gi} S_{2ni}\left(\delta\right) - \Ex[\Omega_{gi} S_{2ni}\left(\delta\right)\mid\Omega_{gi}, Z_i]\right)^2] \leq \sum_{i = 1}^n \Ex\left[\left(\Omega_{gi} S_{2ni}\left(\delta\right)\right)^2\right].
\end{equation*}
Since $\Omega_{gi} S_{2ni}\left(\delta\right)\geq 0$ and $S_{2ni}\left(\delta\right)\leq Z_i\delta/\sqrt{n}$ for all $i = 1,\dots,n$, we can further bound $T_2$ by
\begin{align*}\Ex|T_2|^2 &\leq \max_i\{\Omega_{gi} S_{2ni}\left(\delta\right): 1\leq i \leq n\}\sum_{i = 1}^n \Ex\left[\Omega_{gi} S_{2ni}\left(\delta\right)\right]\\&\leq \frac{1}{\sqrt{n}} \max_i \left|\Omega_{gi}Z_i^{\top} \delta\right| \sum_{i=1}^n \Ex\left[\Omega_{gi} S_{2 n i}\left(\delta\right)\right].
\end{align*}
For the sum we have
$$\sum_{i=1}^n \Ex\left[\Omega_{gi} S_{2 n i}\left(\delta\right)\right] = \Ex\left[\sum_{i=1}^n\Ex\left[\Omega_{gi} S_{2ni}\left(\delta\right)\mid\Omega_{gi},Z_i\right]\right],$$
which can be controlled in the same way as in $T_1$.
Hence, by Assumption~\hyperref[A_variance_bound]{\ref{A_asymptotics1}(iv)}, $T_2$ converges in probability to zero for all $g \in \mathcal{I}$.

Now consider the final summand $T_3$. Since $S_{2ni}\left(\delta\right)\geq 0 $ for all $i = 1,\ldots,n$, the bias due to the estimated weights vanishes as:
$$\sum_{i=1}^n \Ex\left[\left(\widehat{\Omega}_i - \Omega_{gi}\right) S_{2 n i}\left(\delta\right)\right] \leq \sup_{(y,x) \in \mathcal{Y} \times \mathcal{X}} \left|\widehat{\Omega}\left(y,x\right) - \Omega_g\left(y,x\right)\right| \sum_{i=1}^n \Ex\left[S_{2 n i}\left(\delta\right)\right],$$ where the difference $\sup_{(y,x) \in \mathcal{Y} \times \mathcal{X}}|\widehat{\Omega}\left(y,x\right) - \Omega_g\left(y,x\right)| \xrightarrow{p} 0$ uniformly for $g \in \mathcal{I}$ by Assumption~\hyperref[A_sup_Omega]{\ref{A_asymptotics1}(v)}. The second factor converges to a finite limit as in $T_1$:
$$
\begin{aligned}
\sum_{i=1}^n \Ex[S_{2 n i}\left(\delta\right)\mid Z_i] 
& =\left(2 n\right)^{-1} \sum_{i=1}^n f_{Y\mid Z}\left(Z_i^\top \theta_\tau\right) \delta^{\top} Z_i Z_i^{\top} \delta+o_p\left(1\right) \\
& \xrightarrow{p} \frac{1}{2} \delta^{\top} \Ex\big[f_{Y\mid Z}\left(Z_i^\top \theta_\tau\right) Z_i Z_i^{\top}\big] \delta.
\end{aligned}
$$
Thus, $T_3 \xrightarrow{p} 0.$

Combining all three terms we obtain $S_{2n}\left(\delta\right) = T_1 + T_2 + T_3 \xrightarrow{p} \cfrac{1}{2}\delta^{\top} M_{1\tau g} \delta,$ and consequently,
\begin{align*}
S_n\left(\delta\right) &= -\frac{1}{\sqrt{n}} \delta^\top \sum_{i=1}^n M_{0\tau g}(S_i) + \frac{1}{2} \delta^{\top} M_{1\tau g} \delta + o_p(1)\\ &\xrightarrow{d} S_0\left(\delta\right)=-\delta^{\top} \mathcal{Z}_{\tau g}+\frac{1}{2} \delta^{\top} M_{1\tau g} \delta.
\end{align*}
The convexity of the limiting objective function, $S_0\left(\delta\right)$, assures the uniqueness of the minimizer and, consequently, that
$$
\sqrt{n}\left(\widehat{\theta}_\tau -\theta_\tau\right)=\widehat{\delta}=\operatorname{argmin} S_n\left(\delta\right) \xrightarrow{d} \operatorname{argmin} S_0\left(\delta\right) .
$$
(See, \cite{Koenker2005}; \cite{pollard1991}; \cite{knight1998}).
Finally, we conclude the proof by noting that the limiting objective function $S_n(\delta)$ is minimized at \begin{align*}
    \frac{1}{\sqrt{n}}\sum_{i=1}^n\chi_{\tau g}\left(S_i\right) + o_p(1),
\end{align*} recalling the definition of the influence function 
\begin{align*}
    \chi_{\tau,g}(S_i)=M_{1\tau g}^{-1}M_{0\tau g}(S_i).
\end{align*} Therefore, the asymptotic distribution results hold for all $g \in \mathcal{I}$, in particular for $g = g^*$, where $g^*$ minimizes $\Ex\big[\chi_{\tau g}(S)\chi_{\tau g}^\top(S)\big]$, that is, 
$$\sqrt{n}\Big(\widehat{\theta}_{\tau}-\theta_{\tau}\Big) \xrightarrow{d} \mathcal{N}\left(0, \Ex\big[\chi_{\tau g^*}(S)\chi_{\tau g^*}^\top(S)\big]\right).$$
This completes the derivation of the asymptotic distribution of the quantile selection estimator $\widehat{\theta}_\tau$ under the stated regularity conditions. \end{proof}

\renewcommand{\thesection}{A\arabic{section}}
\section{Estimation of Covariance Matrix} \label{appx:cov_mat_est}
In finite samples, the asymptotic covariance matrix 
\(\mathbb{E}\!\left[\chi_{\tau g^*}(S)\chi_{\tau g^*}^\top(S)\right]\)
is estimated using the corresponding sample analogs.  Note that the estimators are based on an empirical analog of sieve projections and are thus identified and do not vary with the inverse selection probability $g$. 
Let 
\[
\widehat{U}_{i\tau} := Y_i - Z_i^\top \widehat{\theta}_\tau,
\]
and let \(\widehat{f}_{Y \mid \Omega_g, Z}(\cdot)\) denote a kernel estimator of the
conditional density of \(Y\) given \((\Omega_g, Z)\).
The first component is estimated by
\begin{align*}
    \widehat{M}_{1\tau}
    &=  
    \mathbb{E}_n\!\left[
        \widehat{\Omega}_i \,
        \widehat{f}_{Y \mid \Omega_g, Z}\!\left(Z_i^\top \widehat{\theta}_\tau\right)
        Z_i Z_i^\top
    \right].
\end{align*}
The second component of the influence function is
\begin{align*}
    \widehat{M}_{0\tau}(S_i)
    &=
    Z_i \widehat{\Omega}_i \, \psi_\tau(\widehat{U}_{i\tau})
    +
    \widehat{T}^\top
        \big(\widehat{H}\widehat{G}^{-1}\widehat{H}^\top\big)^{-1}
        \widehat{H} \widehat{G}^{-1}
        b^K(W_i,X_i)\,\widehat{U}_{J,i}.
\end{align*}
The first-stage quantities are estimated as
\begin{align*}
    \widehat{H}
    &= 
    \mathbb{E}_n\!\left[D_i\,\phi^J(Y_i,X_i)\, b^K(W_i,X_i)^\top\right], \\
    \widehat{G}
    &= 
    \mathbb{E}_n\!\left[b^K(W_i,X_i)\, b^K(W_i,X_i)^\top\right], \\
    \widehat{U}_{J,i}
    &= 
    1 - D_i \, \phi^J(Y_i,X_i)^\top \widehat{\beta}.
\end{align*}
The plug-in covariance estimator of $\mathbb{E}\!\left[\chi_{\tau g^*}(S)\chi_{\tau g^*}^\top(S)\right]$ is then
\[
\widehat{\Sigma}_{\tau}
=
\widehat{M}_{1\tau}^{-1}\mathbb{E}_n\!\left[
    \widehat{M}_{0\tau}(S_i)\,
    \widehat{M}_{0\tau}(S_i)^\top
\right]\widehat{M}_{1\tau}^{-1}.
\]
Consistency of the plug-in estimators follows from the standard
large-sample arguments for quantile regression in \cite{Koenker2005},
together with the rate condition on \(J\) in
Assumption~\hyperref[A_rate_J]{\ref{A_asymptotics1}(i)}.

\newpage
\renewcommand{\thesection}{B}

\section{Outcome variable}\label{appx:outcome_var}
\normalsize

The empirical analysis focuses on the full-time daily wages of German men and women who are active in the labor force, excluding individuals in education, the self-employed, and civil servants. The construction of the wage variable for individuals employed in 2017 follows a multi-stage procedure, largely based on \cite{dauth2020preparing}, with additional modifications tailored to the analysis in this paper.
Table \ref{tab:appx_outcome_var} reports summary statistics for the relevant daily wage variables at each stage of data preparation. All values are conditional on full-time employment. Only the most pertinent wage variables from each step are shown, without repeating intermediate versions.

\setcounter{table}{0}
\renewcommand\thetable{B.\arabic{table}}

\begin{itemize}
    \item \textbf{Stage I} : Raw data -- \texttt{tentgelt}, the original SIAB daily earnings variable (Tagesentgelt / täglicher Leistungssatz).
    \item \textbf{Stage II} : Inflation adjustment -- \texttt{wage\_defl}, the daily wage deflated to 2015 euros using the CPI.
    \item \textbf{Stage III} : Censoring and ceiling adjustment — \texttt{wage}, top-coded at four euros below the contribution assessment ceiling. Also, the censored wage \texttt{ln\_wage\_cens} equals the log wage if uncensored and 0 otherwise.
    \item \textbf{Stage IV} : Trimmed wage -- \texttt{tln\_wage}, obtained after excluding non-German individuals and trimming at the 5th and 95th percentiles.
    \item \textbf{Stage V} : Panel construction — transformation of the dataset into a yearly panel format.
    \item \textbf{Stage VI} : Final sample selection — \texttt{ln\_wage\_2017}, restricted to individuals aged 25–50 in 2017 with a valid instrument available.
\end{itemize}

\clearpage
     \begin{table}[ht]
     \centering
      \caption{Data preparation}
      \label{tab:appx_outcome_var}
    \begin{threeparttable}
    \resizebox{\linewidth}{!}{
    \begin{tabular}{l *{10}{c}}
    \toprule
    &\multicolumn{5}{c}{Men}                               &\multicolumn{5}{c}{Women}     \\
\cmidrule(lr{0.8em}){2-6}
            \cmidrule(lr{0.8em}){7-11} 
    &\multicolumn{1}{c}{Mean}&\multicolumn{1}{c}{Median}&\multicolumn{1}{c}{Max}&\multicolumn{1}{c}{SD}&\multicolumn{1}{c}{Obs.}&\multicolumn{1}{c}{Mean}&\multicolumn{1}{c}{Median}&\multicolumn{1}{c}{Max}&\multicolumn{1}{c}{SD}&\multicolumn{1}{c}{Obs.}\\ 
\cmidrule(lr{0.8em}){2-6}
            \cmidrule(lr{0.8em}){7-11}  \\
\textbf{Stage I}         \\
\cmidrule(lr{0.8em}){1-6}
            \cmidrule(lr{0.8em}){7-11}
tentgelt\_gr &       75.30&       67.00& 208 & 40.19 & 16,394,987 &       58.33&       51.00&  208 & 36.39 & 8,613,552 \\
\cmidrule(lr{0.8em}){1-6}
            \cmidrule(lr{0.8em}){7-11} \\
\textbf{Stage II}          \\
\cmidrule(lr{0.8em}){1-6}
            \cmidrule(lr{0.8em}){7-11}
tentgelt    &       75.30&       67.00& 208 &  40.19 & 16,394,987&       58.33&       51.00&  208 & 36.39 & 8,613,552\\
wage\_defl   &       97.91&       93.02&   203.92 &    42.36 & 16,394,987&       74.77&       70.50& 208 &   38.83 & 8,613,552\\
\cmidrule(lr{0.8em}){1-6}
            \cmidrule(lr{0.8em}){7-11} \\
\textbf{Stage III}        \\            
\cmidrule(lr{0.8em}){1-6}
            \cmidrule(lr{0.8em}){7-11}
wage        &       97.37&       93.02&   203.92 &  41.53 & 16,394,987&      74.61&       70.50&  203.92 &    38.46 & 8,613,552\\
ln\_wage\_cens & 4.42 & 4.48 & 5.32 &  0.48 &  14,497,003 & 4.18 & 4.25 & 5.32 & 0.56 & 8,143,881\\
\cmidrule(lr{0.8em}){1-6}
            \cmidrule(lr{0.8em}){7-11}\\
 \textbf{Stage IV}      \\
\cmidrule(lr{0.8em}){1-6}
            \cmidrule(lr{0.8em}){7-11}
wage        &      100.73&       96.37&  203.92 & 42.54&  13,488,223  &   75.91&       71.93&  203.92 & 38.79 & 7,376,056\\
ln\_wage & 4.53 & 4.57 & 5.32 & 0.49 &  13,324,958 & 4.22 & 4.29 &  5.32 & 0.57 & 7,180,494 \\
ln\_wage\_cens  & 4.45 & 4.52 & 5.32 & 0.47 &  11,751,680 & 4.20 & 4.27 &  5.32 & 0.55 & 6,961,682\\
tln\_wage  & 4.48 & 4.52 & 5.30 & 0.41 &  11,453,354 & 4.17 & 4.25 &  5.05 & 0.47 & 6,333,783\\
\cmidrule(lr{0.8em}){1-6}
            \cmidrule(lr{0.8em}){7-11} \\
\textbf{Stage V}      \\
\cmidrule(lr{0.8em}){1-6}
            \cmidrule(lr{0.8em}){7-11}
wage        &      108.78&   103.48 & 200.75 & 38.48&  9,129,764  &  81.48&       77.93& 200.75 &  37.19 & 4,830,944\\
ln\_wage & 4.63 & 4.64 & 5.30 & 0.39 &  9,088,967 & 4.30 & 4.36 & 5.30 &  0.50 & 4,769,570 \\
ln\_wage\_cens  & 4.55 & 4.58 & 5.30 & 0.36 &  7,784,179 & 4.28 & 4.35 & 5.30 &  0.48 & 4,615,327\\
tln\_wage  & 4.56 & 4.59 & 5.30 & 0.34 &  7,736,565 & 4.24 & 4.31 & 5.05 & 0.43 & 4,231,846\\
\cmidrule(lr{0.8em}){1-6}
            \cmidrule(lr{0.8em}){7-11} \\
\textbf{Stage VI}        \\
\cmidrule(lr{0.8em}){1-6}
            \cmidrule(lr{0.8em}){7-11}
ln\_wage\_2017 & 4.65 & 4.67  & 5.30 & 0.37 &  112,826 & 4.45 & 4.21 & 5.05 & 0.39 & 52,506 \\                           \\
\bottomrule
       \end{tabular}}
       \end{threeparttable}
       \begin{threeparttable}
       \vspace{-1em}
        \begin{tablenotes} \footnotesize
	\item \textit{Notes:}  SIAB-R, 1975 - 2017. Summary statistics of the daily earnings variable at each stage of the data preparation. All values are conditional on being full-time employed. Only relevant wage variables from each step, without repeating everything again, are shown.
    \end{tablenotes}
      \end{threeparttable}
    \end{table}

\newpage
\renewcommand{\thesection}{C}
\section{Estimation of the Distribution Function}\label{appx:cdf}
\setcounter{figure}{0}
\renewcommand\thefigure{C.\arabic{figure}}
Using the estimator $\widehat{g}_n(y,x)$ from the first stage as discussed in Section \ref{ssec:estimation} we can estimate
\begin{equation}
\widehat F_n(y) =  \sum_{i=1, D_i=1}^n \1\{Y_i\leq y\} \widehat \omega_n(Y_i, X_i)
\label{eq:cdfs}
\end{equation}
where the normalized inverse probability estimators are given by
\begin{equation}
\widehat \omega_n (y, x) =  \frac{\widehat g_n(y, x)}{\sum_{i=1, D_i=1}^n \widehat g_n(Y_i, X_i)},
\label{eq:probability_weights}
\end{equation}
which accounts for potential selection bias by weighing the contribution of each observation based on its probability of being observed.

\begin{figure}[ht]
 \centering
 		\caption{Empirical and selection-corrected distributions by gender}
   \label{fig:cdfs_all}
  \begin{threeparttable}
    \centering
      \includegraphics[height = 8cm]{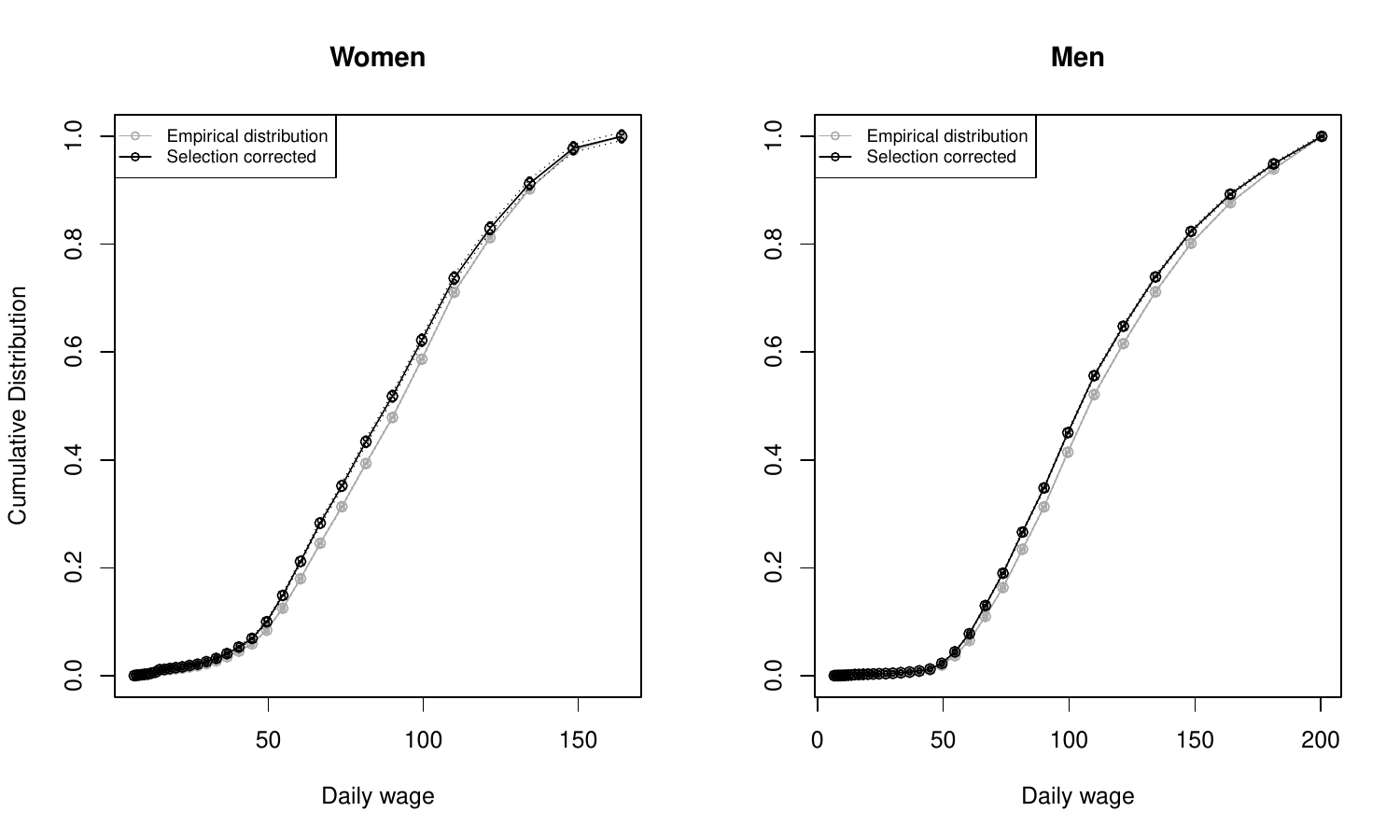}
	\begin{tablenotes} \footnotesize
	   \item \textit{Note: SIAB-R data, 2017 cross-section. The gray lines represent the empirical distributions of observed full-time wages, while the black lines depict the selection-corrected unconditional distributions, estimated using the IV-based selection correction method. All results are shown with 95\% confidence bands.}
	\end{tablenotes}
\end{threeparttable}
\end{figure}

In Figure \ref{fig:cdfs_all}, we illustrate the empirical full-time wage distribution for women (left panel) and men (right panel), alongside the selection-corrected distributions shown in lighter colors, which are estimated using the inverse probability weighting method described in Equation \ref{eq:cdfs}. The controls used in the first stage are detailed in Section \ref{ssec:empirical_model}. The comparison between the two distributions highlights important differences in the wage structure for men and women.

\clearpage
\renewcommand{\thesection}{D}
\section{Simulation Design and Full Results}\label{appx:simulation}
\setcounter{table}{0}
\renewcommand\thetable{D.\arabic{table}}

This section provides a detailed description of the Monte Carlo study introduced in Section \ref{sec:simulation}. We compare its finite-sample performance to existing approaches in the literature. For each scenario, we conduct 1{,}000 Monte Carlo replications with a sample size of $n = 1,000$.

We consider the following model for the outcome:
\[Y^*_i = \beta_0 + \beta_1 W_{i} + \beta_2 X_{i} + \epsilon_i(\tau),\quad i = 1, 2, \ldots, n,\]
where $(W_{i}, X_{i})^\top \sim \mathcal{N}\left(\begin{pmatrix}
2 \\ 1 \end{pmatrix},\begin{pmatrix} 1 & 0.5\\ 0.5 & 1 \end{pmatrix}\right)$. The true parameter vector is set to $(\beta_0, \beta_1, \beta_2) = (1,1,2)$. The error term $\epsilon(\tau)$ is defined as $\epsilon_i(\tau)=\epsilon_i-F_\epsilon^{-1}(\tau)$, where $F_\epsilon(\tau)$ is the $\tau$-quantile of the error distribution. Following closely the simulation design of \cite{zhang_wang}, we consider five different error distributions: (\textbf{A}) $\epsilon_i$ is distributed as $\mathcal{N}(0,1)$; (\textbf{B}) $\epsilon_i$ follows a mixture of normals $0.4\mathcal{N}(0,1.5^2) + 0.6 \mathcal{N}(0,1)$; (\textbf{C}) $\epsilon_i$ follows a $t$-distribution with 3 degrees of freedom, scaled by 0.7; (\textbf{D}) $\epsilon_i$ follows a uniform distribution $\sim \mathcal{U}(-1.5, 1.5)$; (\textbf{E}) $\epsilon_i$ is distributed as $\mathcal{N}(0, (0.5\cdot(1+|X_{i}|))^2)$. Note that settings (\textbf{A})-(\textbf{D}) are homoskedastic errors and (\textbf{E}) is heteroskedastic error.

We generate $\Delta_i$ from the Bernoulli distribution with probability $p(Y^*_i, X_i)$, considering three selection mechanisms:
\begin{itemize}
    \item[\textbf{M1}] $p(Y^*_i, X_i) = 1 / (1 + \exp (- (\alpha_1 + \gamma_1 X_{i})))$ with $(\alpha_1, \gamma_1) = (-0.1, 0.8)$
    \item[\textbf{M2}] $p(Y^*_i, X_i) = 1 / (1 + \exp (- (\alpha_2 + \gamma_2 X_{i} + \xi_2 Y^*_i )))$ with $(\alpha_2, \gamma_2, \xi_2) = (-2.4, 0.6, 0.6)$
    \item[\textbf{M3}] $p(Y^*_i, X_i) = 1 / (1 + \exp (- (\alpha_3 + \gamma_3 \sin^2(X_{i}) + \xi_3 Y^*_i )))$ with $(\alpha_3, \gamma_3, \xi_3)$ $ = (-2.6, 1.2, 0.6)$.
\end{itemize}

The coefficients in the selection models are calibrated to achieve an average missing data rate of approximately 35\%. In M1, the selection probability is independent of the latent outcome $Y^*_i$, reflecting a missing-at-random scenario. In contrast, M2 and M3 introduce dependence on the latent outcome, representing a missing-not-at-random setting. Notably, $W_{i}$ serves as an instrumental variable, influencing the latent outcome but being excluded from the selection equation.

We consider the four estimators as described in Section~\ref{sec:simulation} at the median quantile ($\tau$ = 0.5): the uncorrected estimator, the MAR-assumed correction, the joint estimating equations (JEE) approach of \citet{yu_etal}, and our proposed semiparametric IV estimator. For completeness, we briefly summarize them here. The uncorrected estimator uses only observed outcomes; the MAR-assumed estimator corrects for selection under the MAR assumption via inverse probability weighting; the JEE method employs an IV and specifies a parametric selection model, jointly estimating the quantile and selection equations; and the semiparametric IV estimator proposed in this paper avoids parametric restrictions on the selection mechanism.

We assess the finite-sample estimation performance in terms of the mean biases and the RMSE. Additionally, we report the coverage probabilities of the 95\% confidence intervals along with their average lengths. The uncorrected and MAR-assumed correction methods are implemented using the \texttt{quantreg} package in R, while the JEE method follows the implementation described in \cite{yu_etal}. Tables \ref{tab:settingA}-\ref{tab:settingE} present the simulation results.
\newpage
\FloatBarrier
\begin{table}[ht!]\footnotesize
\caption{(Setting A) Simulation results by method with $(\tau, n)= (0.5, 1000)$ and 35\% missing in the outcome}
\centering
\resizebox{0.99\linewidth}{!}{%
\begin{tabular}[t]{lrrr|rrr|rrr}
\toprule
\multirow{2}{*}{Method} & \multicolumn{3}{c|}{M1} & \multicolumn{3}{c|}{M2} & \multicolumn{3}{c}{M3}\\
& \multicolumn{1}{c}{$\beta_0$} & \multicolumn{1}{c}{$\beta_1$} & \multicolumn{1}{c|}{$\beta_2$} & \multicolumn{1}{c}{$\beta_0$} & \multicolumn{1}{c}{$\beta_1$} & \multicolumn{1}{c|}{$\beta_2$} & \multicolumn{1}{c}{$\beta_0$} & \multicolumn{1}{c}{$\beta_1$} & \multicolumn{1}{c}{$\beta_2$} \\
\midrule
& \multicolumn{9}{c}{Mean biases} \\
\cmidrule{2-10}
Uncorrected & 0.000 & -0.002 & 0.003 & 0.342 & -0.035 & -0.106 & 0.351 & -0.039 & -0.101\\
MAR-assumed &0.002 &-0.002 & 0.001 & 0.391 & -0.040 & -0.128 & 0.386 & -0.045 & -0.112\\
JEE & 0.002 & -0.001 & 0.000 & 0.051 & -0.009 & -0.012 & 0.055 & -0.011 & -0.008\\
Semiparametric IV & -0.047 & 0.010 & 0.022 & 0.048 & 0.007 & -0.027 & 0.082 & 0.001 & -0.045\\
\addlinespace
\cmidrule{2-10}
& \multicolumn{9}{c}{RMSE} \\
\cmidrule{2-10}
Uncorrected & 0.122 & 0.057 & 0.057 & 0.365 & 0.064 & 0.122 & 0.376 & 0.068 & 0.118\\
MAR-assumed & 0.126 & 0.059 & 0.061 & 0.439 & 0.085 & 0.175 & 0.444 & 0.092 & 0.169\\
JEE & 0.134 & 0.06 & 0.063 & 0.273 & 0.083 & 0.134 & 0.318 & 0.102 & 0.137\\
Semiparametric IV & 0.137 & 0.059 & 0.068 & 0.156 & 0.062 & 0.079 & 0.170 & 0.062 & 0.086\\
\addlinespace
\cmidrule{2-10}
& \multicolumn{9}{c}{CI lengths} \\
\cmidrule{2-10}
Uncorrected & 0.464 & 0.223 & 0.233 & 0.523 & 0.219 & 0.241 & 0.519 & 0.220 & 0.237\\
MAR-assumed & 0.485 & 0.230 & 0.250 & 0.612 & 0.226 & 0.326 & 0.519 & 0.198 & 0.183\\
JEE & 0.527 & 0.240 & 0.263 & 0.934 & 0.314 & 0.418 & 0.972 & 0.341 & 0.444\\
Semiparametric IV & 0.696 & 0.342 & 0.317 & 0.683 & 0.325 & 0.272 & 0.735 & 0.347 & 0.282 \\
\addlinespace
\cmidrule{2-10}
& \multicolumn{9}{c}{Coverage probabilities} \\
\cmidrule{2-10}
Uncorrected & 0.939 & 0.956 & 0.963 & 0.256 & 0.917 & 0.600 & 0.252 & 0.899 & 0.620\\
MAR-assumed & 0.923 & 0.948 & 0.955 & 0.377 & 0.892 & 0.651 & 0.431 & 0.883 & 0.702\\
JEE & 0.941 & 0.949 & 0.944 & 0.886 & 0.935 & 0.884 & 0.853 & 0.900 & 0.872\\
Semiparametric IV & 0.979 & 0.995 & 0.968 & 0.948 & 0.959 & 0.955 & 0.958 & 0.989 & 0.877\\
\bottomrule
\bottomrule
\end{tabular}}
\label{tab:settingA}

\end{table}

\newpage
\FloatBarrier
\begin{table}[h!]\footnotesize
\caption{(Setting B) Simulation results by method with $(\tau, n)= (0.5, 1000)$ and 35\% missing in the outcome}
\centering
\resizebox{0.99\linewidth}{!}{%
\begin{tabular}[t]{lrrr|rrr|rrr}
\toprule
\multirow{2}{*}{Method} & \multicolumn{3}{c|}{M1} & \multicolumn{3}{c|}{M2} & \multicolumn{3}{c}{M3}\\
& \multicolumn{1}{c}{$\beta_0$} & \multicolumn{1}{c}{$\beta_1$} & \multicolumn{1}{c|}{$\beta_2$} & \multicolumn{1}{c}{$\beta_0$} & \multicolumn{1}{c}{$\beta_1$} & \multicolumn{1}{c|}{$\beta_2$} & \multicolumn{1}{c}{$\beta_0$} & \multicolumn{1}{c}{$\beta_1$} & \multicolumn{1}{c}{$\beta_2$} \\
\midrule
& \multicolumn{9}{c}{Mean biases} \\
\cmidrule{2-10}
Uncorrected & 0.006 & -0.003 & 0.000 & 0.246 & -0.022 & -0.079 & 0.257 & -0.026 & -0.078\\
MAR-assumed & 0.005 & -0.003 & -0.001 & 0.287 & -0.026 & -0.099 & 0.294 & -0.032 & -0.092\\
JEE & 0.004 & -0.002 & -0.002 & 0.020 & -0.001 & -0.008 & 0.044 & -0.009 & -0.010\\
Semiparametric IV & -0.029 & 0.005 & 0.016 & 0.044 & 0.002 & -0.019 & 0.055 & 0.003 & -0.034\\
\addlinespace
\cmidrule{2-10}
& \multicolumn{9}{c}{RMSE} \\
\cmidrule{2-10}
Uncorrected & 0.101 & 0.050 & 0.051 & 0.272 & 0.054 & 0.095 & 0.280 & 0.055 & 0.093\\
MAR-assumed & 0.106 & 0.051 & 0.056 & 0.338 & 0.073 & 0.136 & 0.352 & 0.077 & 0.138\\
JEE & 0.111 & 0.052 & 0.057 & 0.233 & 0.076 & 0.103 & 0.269 & 0.084 & 0.118\\
Semiparametric IV & 0.110 & 0.050 & 0.056 & 0.135 & 0.052 & 0.063 & 0.139 & 0.053 & 0.068\\
\addlinespace
\cmidrule{2-10}
& \multicolumn{9}{c}{CI lengths} \\
\cmidrule{2-10}
Uncorrected & 0.393 & 0.189 & 0.197 & 0.446 & 0.187 & 0.206 & 0.446 & 0.189 & 0.203\\
MAR-assumed & 0.408 & 0.194 & 0.211 & 0.672 & 0.268 & 0.414 & 0.710 & 0.275 & 0.377\\
JEE & 0.437 & 0.204 & 0.222 & 0.758 & 0.264 & 0.344 & 0.803 & 0.291 & 0.376\\
Semiparametric IV & 0.622 & 0.308 & 0.287 & 0.599 & 0.280 & 0.237 & 0.642 & 0.304 & 0.245\\
\addlinespace
\cmidrule{2-10}
& \multicolumn{9}{c}{Coverage probabilities} \\
\cmidrule{2-10}
Uncorrected & 0.933 & 0.936 & 0.946 & 0.415 & 0.918 & 0.703 & 0.374 & 0.922 & 0.666\\
MAR-assumed & 0.941 & 0.943 & 0.942 & 0.509 & 0.896 & 0.723 & 0.517 & 0.895 & 0.744\\
JEE & 0.939 & 0.946 & 0.937 & 0.887 & 0.921 & 0.906 & 0.842 & 0.920 & 0.899\\
Semiparametric IV & 0.985 & 0.996 & 0.981 & 0.961 & 0.992 & 0.949 & 0.968 & 0.992 & 0.917\\
\bottomrule
\bottomrule
\end{tabular}}
\end{table}

\newpage
\FloatBarrier
\begin{table}[h!]\footnotesize
\caption{(Setting D) Simulation results by method with $(\tau, n)= (0.5, 1000)$ and 35\% missing in the outcome}
\centering
\resizebox{0.99\linewidth}{!}{%
\begin{tabular}[t]{lrrr|rrr|rrr}
\toprule
\multirow{2}{*}{Method} & \multicolumn{3}{c|}{M1} & \multicolumn{3}{c|}{M2} & \multicolumn{3}{c}{M3}\\
& \multicolumn{1}{c}{$\beta_0$} & \multicolumn{1}{c}{$\beta_1$} & \multicolumn{1}{c|}{$\beta_2$} & \multicolumn{1}{c}{$\beta_0$} & \multicolumn{1}{c}{$\beta_1$} & \multicolumn{1}{c|}{$\beta_2$} & \multicolumn{1}{c}{$\beta_0$} & \multicolumn{1}{c}{$\beta_1$} & \multicolumn{1}{c}{$\beta_2$} \\
\midrule
& \multicolumn{9}{c}{Mean biases} \\
\cmidrule{2-10}
Uncorrected & 0.004 & -0.002 & -0.001 & 0.381 & -0.035 & -0.123 & 0.377 & -0.037 & -0.113\\
MAR-assumed & 0.006 & -0.002 & -0.001 & 0.434 & -0.039 & -0.151 & 0.403 & -0.043 & -0.118\\
JEE & 0.003 & -0.002 & 0.000 & 0.059 & -0.004 & -0.025 & 0.054 & -0.009 & -0.013\\
Semiparametric IV & -0.045 & 0.012 & 0.023 & 0.062 & 0.007 & -0.033 & 0.079 & 0.004 & -0.049\\
\addlinespace
\cmidrule{2-10}
& \multicolumn{9}{c}{RMSE} \\
\cmidrule{2-10}
Uncorrected & 0.135 & 0.065 & 0.071 & 0.409 & 0.073 & 0.140 & 0.405 & 0.074 & 0.134\\
MAR-assumed & 0.144 & 0.069 & 0.077 & 0.487 & 0.093 & 0.192 & 0.467 & 0.101 & 0.178\\
JEE & 0.152 & 0.069 & 0.080 & 0.318 & 0.101 & 0.143 & 0.345 & 0.112 & 0.155\\
Semiparametric IV & 0.160 & 0.071 & 0.078 & 0.192 & 0.077 & 0.088 & 0.194 & 0.071 & 0.098\\
\addlinespace
\cmidrule{2-10}
& \multicolumn{9}{c}{CI lengths} \\
\cmidrule{2-10}
Uncorrected & 0.543 & 0.262 & 0.274 & 0.604 & 0.255 & 0.281 & 0.603 & 0.257 & 0.277\\
MAR-assumed & 0.568 & 0.271 & 0.294 & 0.855 & 0.326 & 0.480 & 0.850 & 0.351 & 0.472\\
JEE & 0.621 & 0.285 & 0.311 & 1.005 & 0.355 & 0.448 & 1.021 & 0.379 & 0.465\\
Semiparametric IV & 0.661 & 0.328 & 0.301 & 0.722 & 0.336 & 0.284 & 0.781 & 0.361 & 0.298\\
\addlinespace
\cmidrule{2-10}
& \multicolumn{9}{c}{Coverage probabilities} \\
\cmidrule{2-10}
Uncorrected & 0.944 & 0.962 & 0.936 & 0.289 & 0.924 & 0.597 & 0.294 & 0.922 & 0.633\\
MAR-assumed & 0.938 & 0.951 & 0.934 & 0.383 & 0.900 & 0.633 & 0.484 & 0.888 & 0.697 \\
JEE & 0.935 & 0.953 & 0.923 & 0.863 & 0.924 & 0.866 & 0.817 & 0.906 & 0.828 \\
Semiparametric IV & 0.952 & 0.965 & 0.940 & 0.922 & 0.960 & 0.874 & 0.935 & 0.979 & 0.847 \\ 
\bottomrule
\bottomrule
\end{tabular}}
\end{table}

\newpage
\FloatBarrier
\begin{table}[h!]\footnotesize
\caption{(Setting E) Simulation results by method with $(\tau, n)= (0.5, 1000)$ and 35\% missing in the outcome}
\centering
\resizebox{0.99\linewidth}{!}{%
\begin{tabular}[t]{lrrr|rrr|rrr}
\toprule
\multirow{2}{*}{Method} & \multicolumn{3}{c|}{M1} & \multicolumn{3}{c|}{M2} & \multicolumn{3}{c}{M3}\\
& \multicolumn{1}{c}{$\beta_0$} & \multicolumn{1}{c}{$\beta_1$} & \multicolumn{1}{c|}{$\beta_2$} & \multicolumn{1}{c}{$\beta_0$} & \multicolumn{1}{c}{$\beta_1$} & \multicolumn{1}{c|}{$\beta_2$} & \multicolumn{1}{c}{$\beta_0$} & \multicolumn{1}{c}{$\beta_1$} & \multicolumn{1}{c}{$\beta_2$} \\
\midrule
& \multicolumn{9}{c}{Mean biases} \\
\cmidrule{2-10}
Uncorrected & 0.007 & -0.004 & 0.002 & 0.231 & -0.037 & -0.033 & 0.247 & -0.045 & -0.021\\
MAR-assumed & 0.006 & -0.004 & 0.002 & 0.279 & -0.038 & -0.075 & 0.311 & -0.044 & -0.080\\ 
JEE & 0.006 & -0.003 & 0.001 & 0.061 & -0.011 & -0.016 & 0.068 & -0.016 & -0.010\\ 
Semiparametric IV & -0.032 & 0.007 & 0.037 & -0.052 & 0.024 & 0.045 & -0.016 & 0.018 & 0.001\\
\addlinespace
\cmidrule{2-10}
& \multicolumn{9}{c}{RMSE} \\
\cmidrule{2-10}
Uncorrected & 0.108 & 0.059 & 0.069 & 0.265 & 0.070 & 0.079 & 0.284 & 0.078 & 0.076 \\
MAR-assumed & 0.109 & 0.058 & 0.070 & 0.349 & 0.086 & 0.155 & 0.381 & 0.096 & 0.161 \\ 
JEE & 0.113 & 0.058 & 0.070 & 0.260 & 0.086 & 0.125 & 0.287 & 0.094 & 0.146 \\
Semiparametric IV & 0.118 & 0.058 & 0.081 & 0.155 & 0.067 & 0.106 & 0.151 & 0.069 & 0.092\\
\addlinespace
\cmidrule{2-10}
& \multicolumn{9}{c}{CI lengths} \\
\cmidrule{2-10}
Uncorrected & 0.370 & 0.214 & 0.211 & 0.479 & 0.232 & 0.246 & 0.464 & 0.231 & 0.222\\ 
MAR-assumed & 0.370 & 0.207 & 0.201 & 0.768 & 0.288 & 0.468 & 0.687 & 0.296 & 0.356\\ 
JEE & 0.452 & 0.230 & 0.286 & 0.764 & 0.296 & 0.398 & 0.866 & 0.325 & 0.453\\ 
Semiparametric IV & 0.698 & 0.398 & 0.337 & 0.717 & 0.410 & 0.294 & 0.767 & 0.419 & 0.308\\
\addlinespace
\cmidrule{2-10}
& \multicolumn{9}{c}{Coverage probabilities} \\
\cmidrule{2-10}
Uncorrected & 0.901 & 0.928 & 0.862 & 0.527 & 0.900 & 0.867 & 0.466 & 0.855 & 0.856\\
MAR-assumed & 0.891 & 0.927 & 0.844 & 0.530 & 0.888 & 0.743 & 0.525 & 0.864 & 0.714\\
JEE & 0.946 & 0.941 & 0.944 & 0.889 & 0.947 & 0.915 & 0.859 & 0.911 & 0.885\\
Semiparametric IV & 0.995 & 0.998 & 0.956 & 0.975 & 0.989 & 0.852 & 0.977 & 0.991 & 0.900 \\
\bottomrule
\bottomrule
\end{tabular}}
\label{tab:settingE}

\end{table}
\clearpage
\newpage
\FloatBarrier
\renewcommand{\thesection}{E}
\section{Robustness}\label{appx:robustness}

This section presents several robustness checks. In these analyses, we impose stricter requirements on the lag structure of the instrument. In the main specification, the initial wage instrument includes wages observed up to 2015, using full-time wages where available and part-time wages when no full-time record exists. The robustness checks restrict the instrument in two alternative ways: (i) by using only wages from 2011 or earlier (i.e., at least five years before the outcome period) as valid instruments, and (ii) by limiting the instrument to the earliest observed full-time wage only.

\subsection{Instrument based on wage history from 2011 or prior}

Figures~\ref{fig:uncond_quantiles2011} and \ref{fig:cond_quantiles2011} replicate Figures~\ref{fig:quantiles_uncond2015} and \ref{fig:plot_by_edu}, respectively. The estimated quantile patterns remain virtually unchanged, with effect sizes and gender differences closely mirroring those in the main specification. This confirms that the results are robust to using earlier wage information as the instrument.

\setcounter{figure}{0}
\renewcommand\thefigure{E.\arabic{figure}}
\begin{figure}[htbp]
    \caption{Quantile Selection Effects (Unconditional) -- only IV up to 2011}
    \label{fig:uncond_quantiles2011}
  \begin{threeparttable}
    \centering
    \includegraphics[width=1\linewidth]{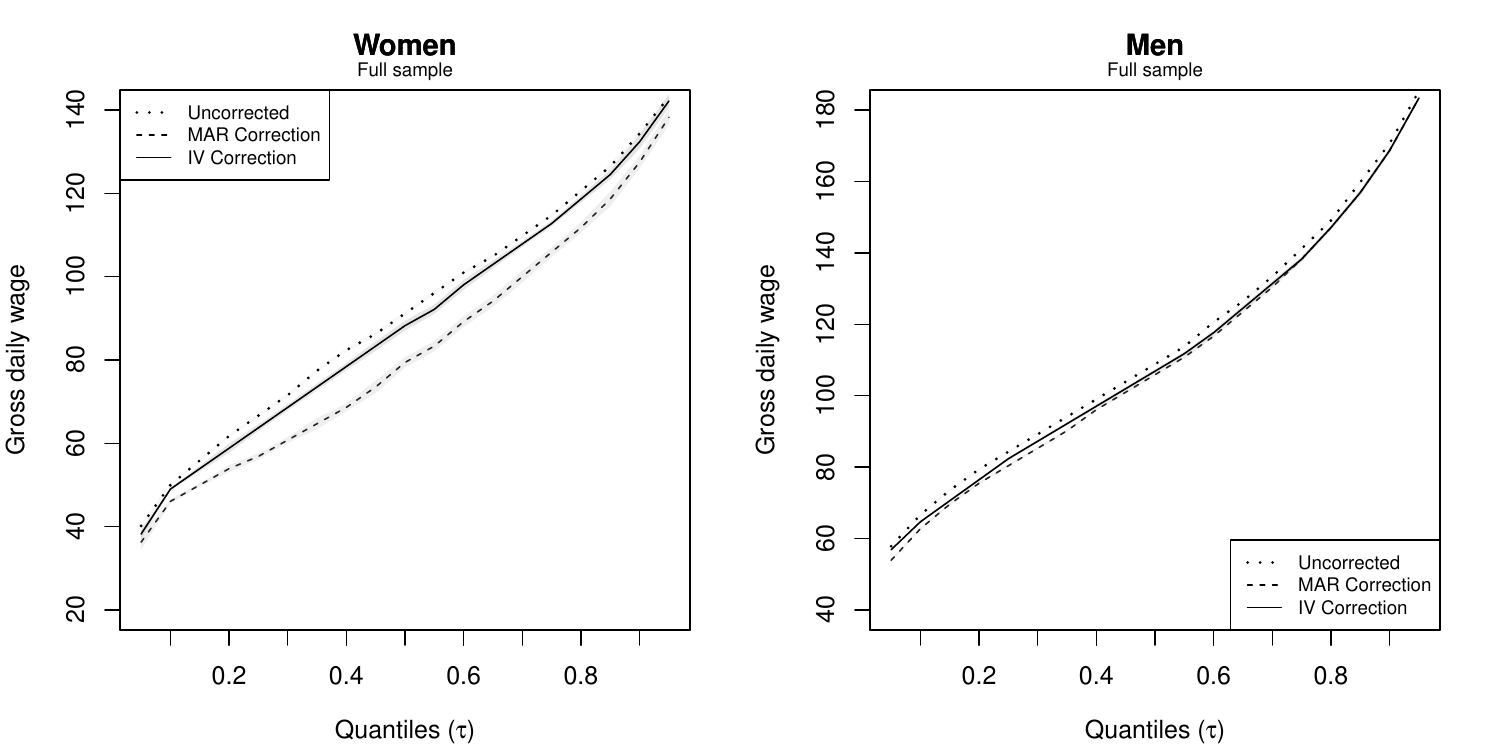}
    \begin{tablenotes} \footnotesize
        \item \textit{Note: SIAB-R data, 2017. Curves display wage point estimates from unconditional quantile regressions across quantiles, connected by lines for visualization: uncorrected (dotted), MAR correction (dashed), and IV correction (solid). The IV correction uses initial log wage as an instrument and controls for education, age, workplace region, total experience, and earliest job difficulty in the first two stages. The MAR correction uses the same controls except earliest job difficulty. The final-stage regressions include no covariates for all methods. Shaded areas represent 95\% pointwise confidence intervals.}
    \end{tablenotes}
    \end{threeparttable}
\end{figure}

\begin{figure}[htbp]
\caption{Conditional wage quantiles by education groups -- only IV up to 2011}
\label{fig:cond_quantiles2011}
  \begin{threeparttable}

	\begin{subfigure}{1\textwidth}
 \centering
 \vspace{-0.5em}
 		\caption{Low Education}
            \includegraphics[width=.87\linewidth]{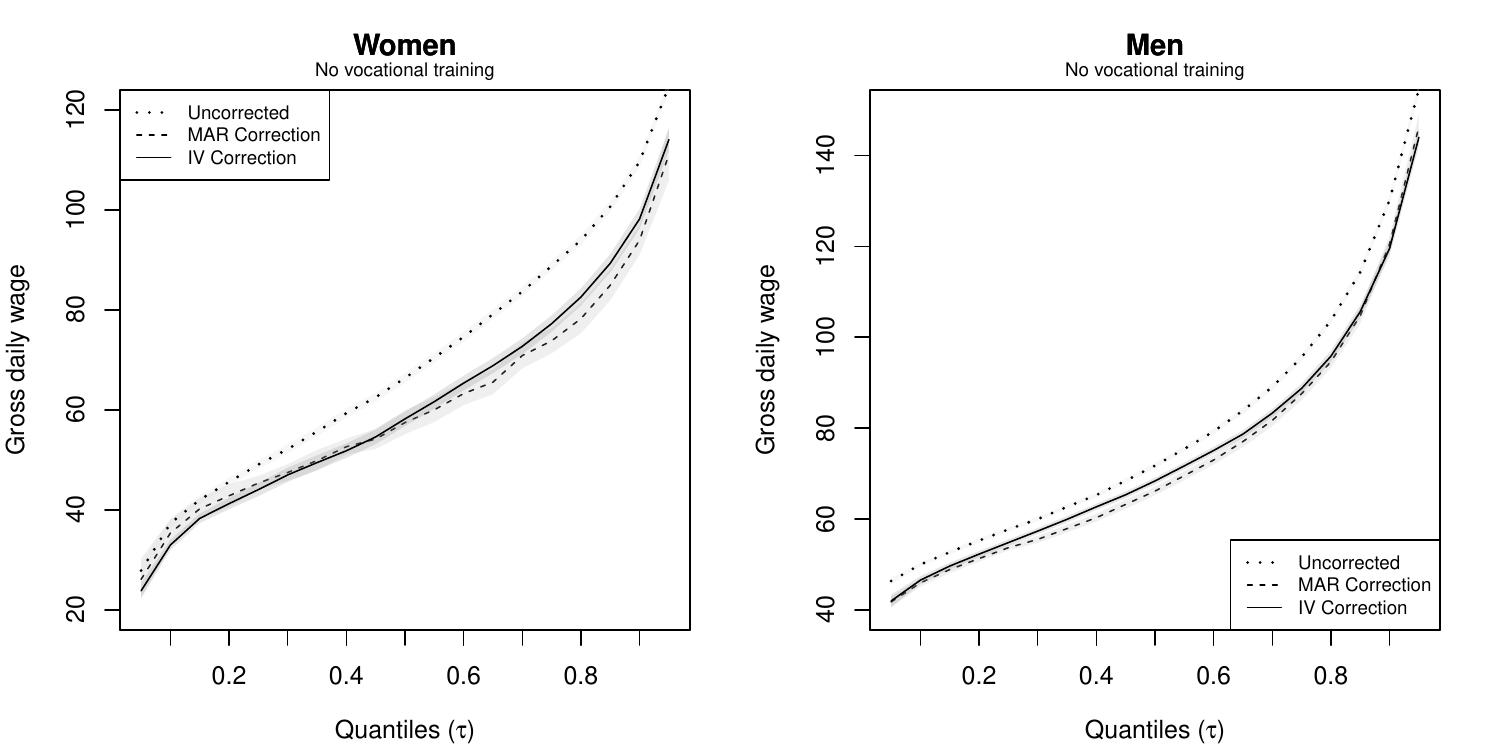}
	\end{subfigure}
 \vspace{-0.3em}
	\begin{subfigure}{1\textwidth}
		\centering
        \vspace{-0.5em}
		\caption{Middle Education}

		\includegraphics[width=.87\linewidth]{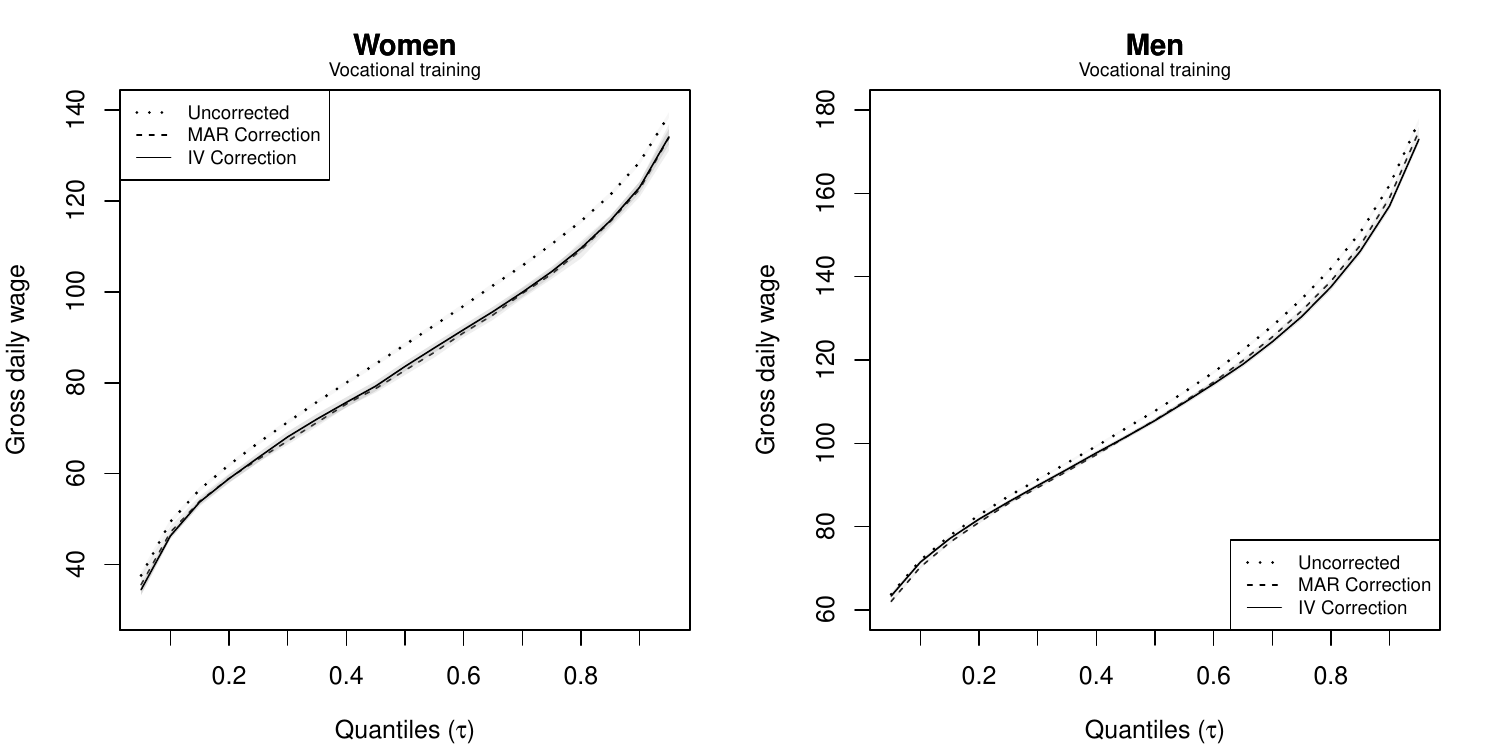}
	\end{subfigure}
\vspace{-0.3em}
 	\begin{subfigure}{1\textwidth}
		\centering
        \vspace{-0.5em}
		\caption{High Education}
		\includegraphics[width=.87\linewidth]{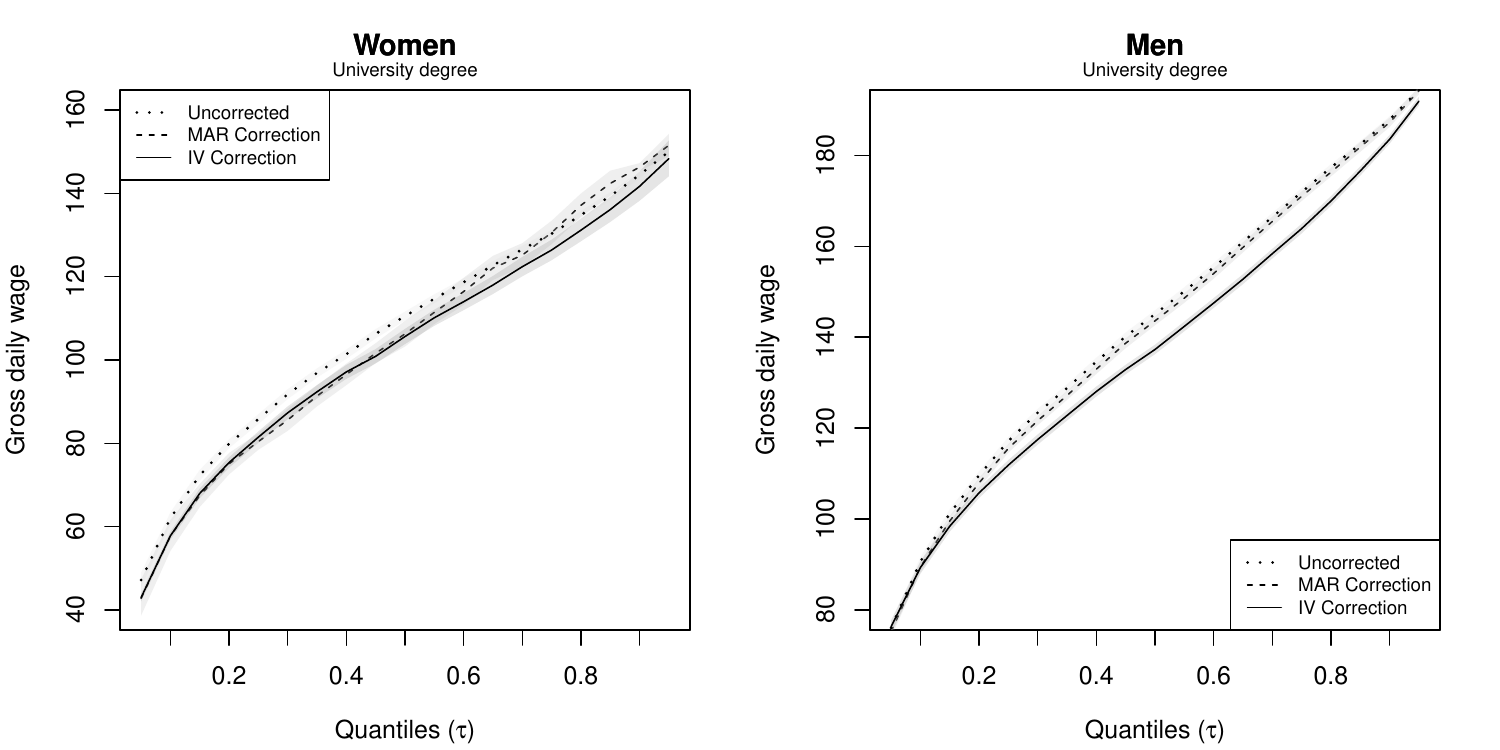}
	\end{subfigure}
    \vspace{-1em}
	\begin{tablenotes}[flushleft] \footnotesize
	    \item \textit{Note: SIAB-R data, 2017. Curves show wage point estimates from unconditional quantile regressions across quantiles, connected by lines for visualization: uncorrected (dotted), MAR correction (dashed), and IV correction (solid). The IV correction uses initial log wage as an instrument and controls for education, age, workplace region, total experience, and earliest job difficulty. The MAR correction uses the same controls, excluding the instrument and earliest job difficulty. The uncorrected estimates include the same controls as in MAR. Shaded areas represent 95\% pointwise confidence intervals.}
	\end{tablenotes}
\end{threeparttable}
\end{figure}

\clearpage
\subsection{Instrument based on only full-time wage history}
Figures~\ref{fig:uncond_quantilesFT} and \ref{fig:cond_quantilesFT} replicate Figures~\ref{fig:quantiles_uncond2015} and \ref{fig:plot_by_edu}, respectively. The estimated quantile patterns remain virtually unchanged, with effect sizes and gender differences closely mirroring those in the main specification. This confirms that the results are robust to using only full-time wage information as the instrument.

\setcounter{figure}{2}
\renewcommand\thefigure{E.\arabic{figure}}
\begin{figure}[htbp]
    \caption{Quantile Selection Effects (Unconditional) -- only full-time instruments}
    \label{fig:uncond_quantilesFT}
  \begin{threeparttable}
    \centering
    \includegraphics[width=1\linewidth]{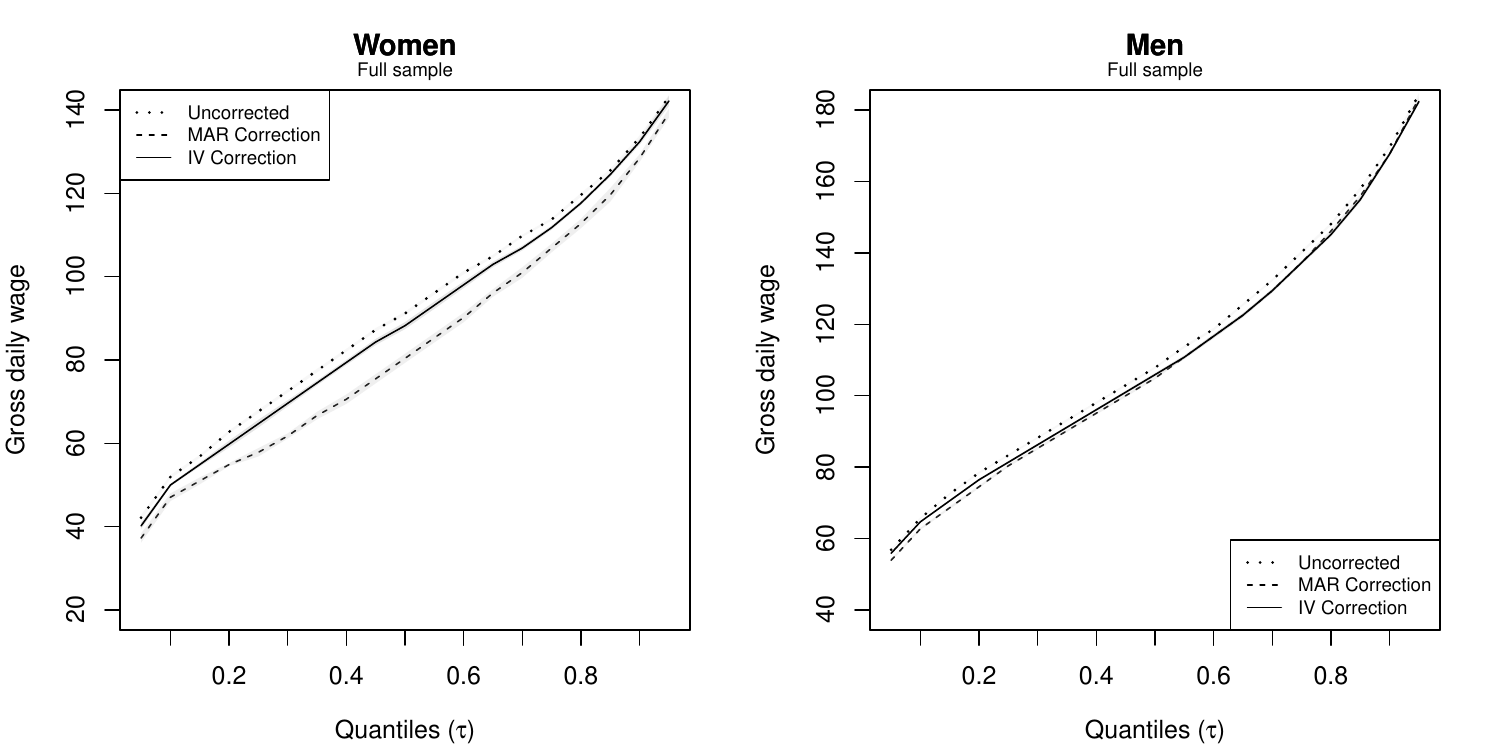}
    \begin{tablenotes} \footnotesize
        \item \textit{Note: SIAB-R data, 2017. Curves display wage point estimates from unconditional quantile regressions across quantiles, connected by lines for visualization: uncorrected (dotted), MAR correction (dashed), and IV correction (solid). The IV correction uses initial log wage as an instrument and controls for education, age, workplace region, total experience, and earliest job difficulty in the first two stages. The MAR correction uses the same controls except earliest job difficulty. The final-stage regressions include no covariates for all methods. Shaded areas represent 95\% pointwise confidence intervals.}
    \end{tablenotes}
    \end{threeparttable}
\end{figure}

\begin{figure}[htbp]
\caption{Conditional wage quantiles by education -- only full-time instruments}
\label{fig:cond_quantilesFT}
  \begin{threeparttable}

	\begin{subfigure}{1\textwidth}
 \centering
 \vspace{-0.5em}
 		\caption{Low Education}
            \includegraphics[width=.87\linewidth]{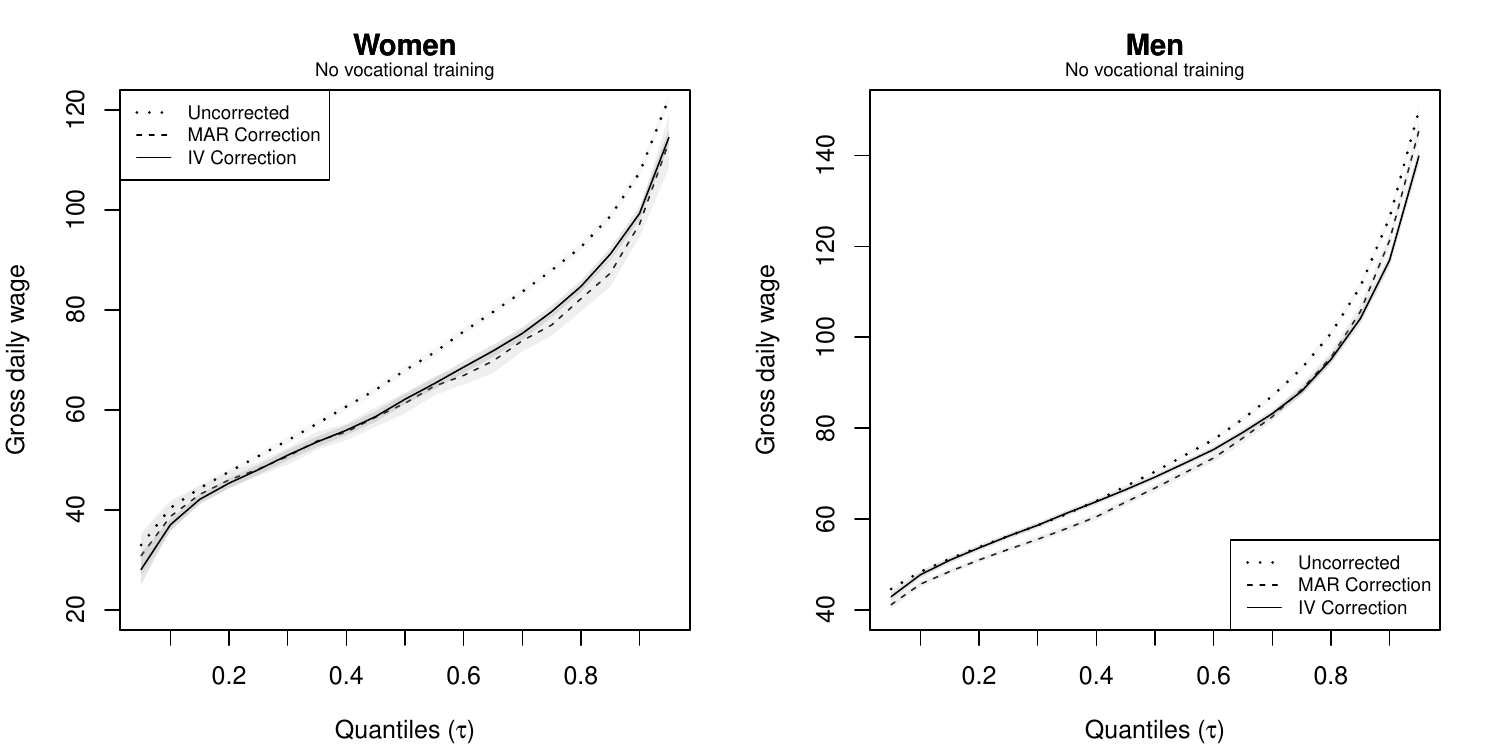}
	\end{subfigure}
 \vspace{-0.3em}
	\begin{subfigure}{1\textwidth}
		\centering
        \vspace{-0.5em}
		\caption{Middle Education}

		\includegraphics[width=.87\linewidth]{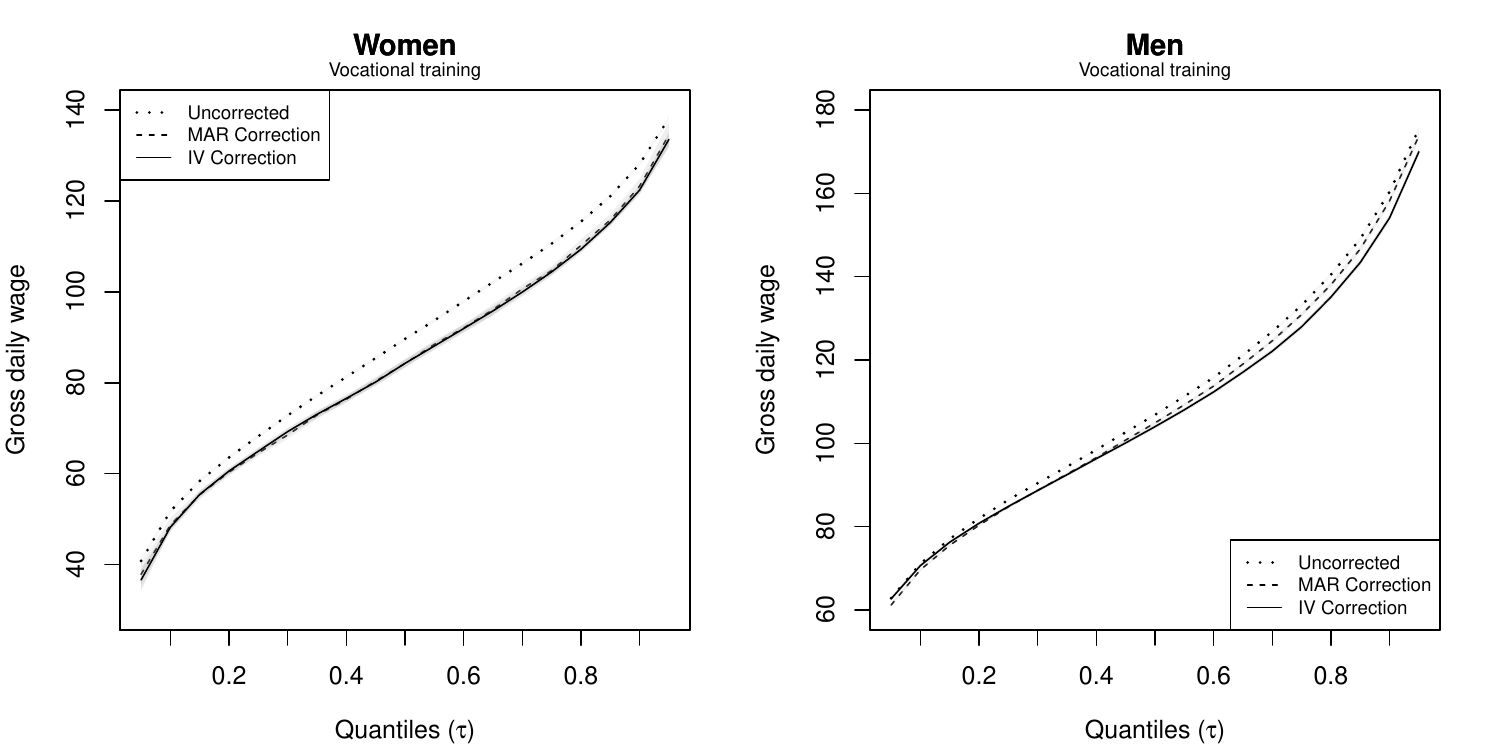}
	\end{subfigure}
\vspace{-0.3em}
 	\begin{subfigure}{1\textwidth}
		\centering
        \vspace{-0.5em}
		\caption{High Education}
		\includegraphics[width=.87\linewidth]{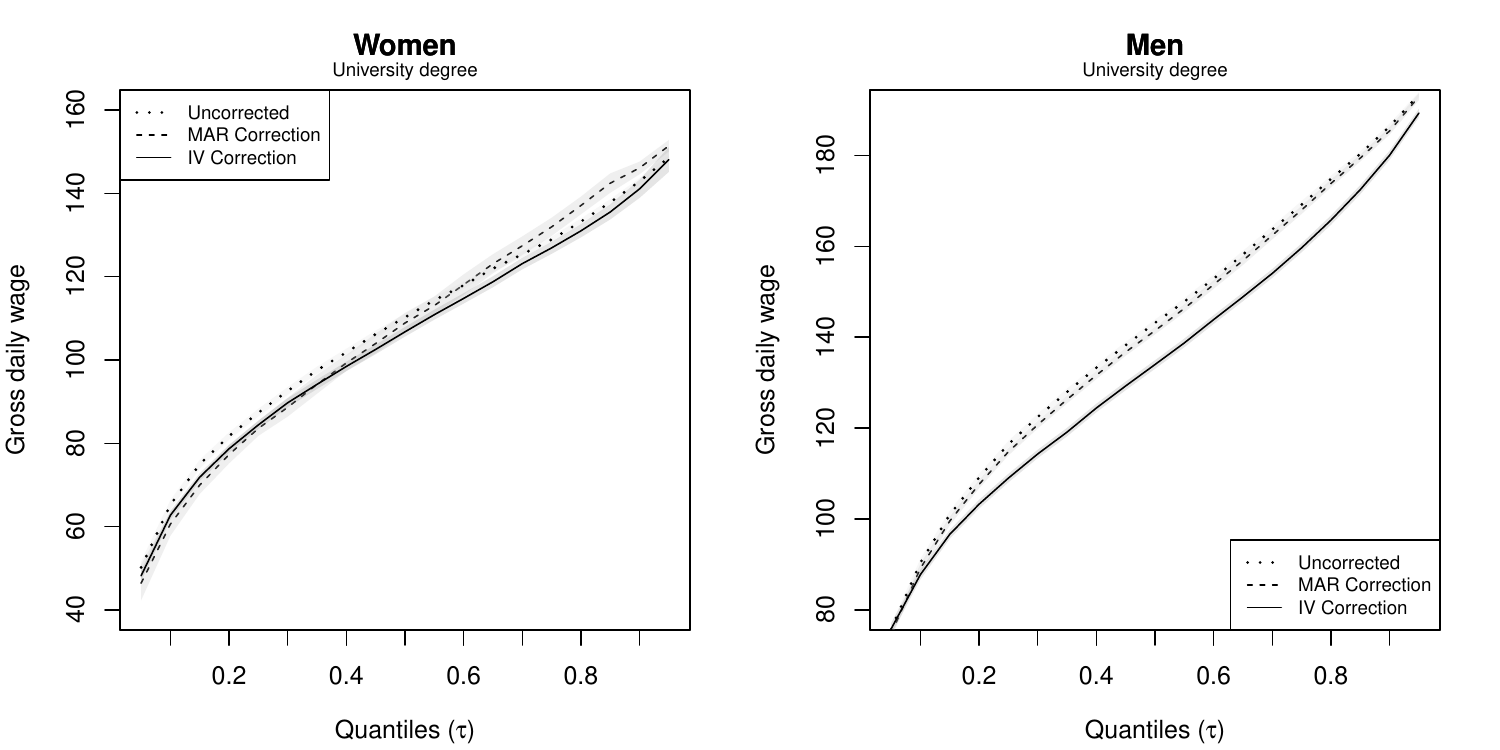}
	\end{subfigure}
    \vspace{-1em}
	\begin{tablenotes}[flushleft] \footnotesize
	    \item \textit{Note: SIAB-R data, 2017. Curves show wage point estimates from unconditional quantile regressions across quantiles, connected by lines for visualization: uncorrected (dotted), MAR correction (dashed), and IV correction (solid). The IV correction uses initial log wage as an instrument and controls for education, age, workplace region, total experience, and earliest job difficulty. The MAR correction uses the same controls, excluding the instrument and earliest job difficulty. The uncorrected estimates include the same controls as in MAR. Shaded areas represent 95\% pointwise confidence intervals.}
	\end{tablenotes}
\end{threeparttable}
\end{figure}

\end{document}